\documentclass[mathleft,fleqn]{an}

\usepackage[square,sort]{natbib}
\usepackage{graphicx}
\usepackage{amstext}
\usepackage{amssymb}
\usepackage{lmodern}
\usepackage{hyperref}
\usepackage{multicol,lipsum,environ}
\usepackage{color}
\usepackage{soul}
\newcommand{\Lim}[1]{\raisebox{0.5ex}{\scalebox{0.8}{$\displaystyle \lim_{#1}\;$}}}

\newcommand{\be}{\begin{equation}}
\newcommand{\ee}{\end{equation}}
\newcommand{\ba}{\begin{eqnarray}}
\newcommand{\ea}{\end{eqnarray}}
\newcommand{\bc}{\begin{center}}
\newcommand{\ec}{\end{center}}

\newcommand{\bL}{\hat \emph{L}}
\newcommand{\bR}{\hat \emph{R}}
\newcommand{\bx}{\mathbf{x}}
\newcommand{\bB}{\mathbf{B}}
\newcommand{\bk}{\mathbf{k}}

\newcommand{\hhh}{}

\newcommand{\hsix}{}

\usepackage{graphicx}
\usepackage[varg]{txfonts}
\begin{document}

\title{On the possibility of helicity oscillations in the saturation of the Tayler instability}
\author{Alfio Bonanno\inst{1}  \and 
Filippo Guarnieri\inst{2}\footnote{\href{mailto:filippo.guarnieri@roma1.infn.it}{filippo.guarnieri@roma1.infn.it}}}
\titlerunning{On helicity oscillations in Tayler instability}
\authorrunning{Alfio Bonanno \& Filippo Guarnieri}
\institute{
INAF - Catania Astrophysical Observatory
\and 
Nordic Institute for Theoretical Physics (NORDITA), Stockholm, Sweden}

\received{12 Nov 2016}
\accepted{16 Feb 2017}
\publonline{later}


\abstract{%
Recent numerical results of current-driven instabilities at low magnetic Prandtl number and high Hartmann number support the possibility of a saturation state characterized by helicity oscillations. We investigate the underlying mechanism by analyzing this possibility using an higher-order Landau-Ginzburg effective Lagrangian for the weakly non-linear amplitude dynamics, where the magnetic and velocity perturbations are linearly dependent. We find that, if the mirror symmetry between left- and right-handed modes is spontaneously broken, it is impossible to achieve an oscillating helical state. We argue that the result is likely to hold also adding higher-order terms and in the presence of an explicit symmetry breaking. We conclude that an oscillating saturating state for the Tayler instability is unlikely to depend on the interaction of chiral modes.}
\maketitle

\section{Introduction}
The Tayler instability is the instability of a toroidal field in a stably stratified medium due to the presence of an electric current along the axis of symmetry  \citep{tayler73}. 
In its simplest realization, in cylindrical geometry with axial symmetry, it can be shown that a purely toroidal field $B_\phi(r)$, where $r$ is the cylindrical radius, is stable against axisymmetric perturbations if it satisfies the condition $d(B_\phi/r)/dr<0$ and to non-axisymmetric perturbations if $d(r\, B_\phi^2)/dr \leq 0$.

In recent times this problem has attracted a considerable amount of analytical and numerical investigations \citep{bour11,bodo13,jouve15,ibanez15}, 
as well as experimental verification \citep{seil12}, mainly for its relevance for our understanding of magnetism in stellar radiative regions \citep{bra15,kit08} as well as for its implications in the tachocline stability problem and the solar dynamo \citep{kit07,bo13}. 

In particular it was noticed in \cite{bour08} that the unstable modes are helical and, in the presence of a vertical component $B_z$, also invariant under the symmetry transformation $P_{LR}{}^z: (m,\epsilon)\rightarrow  (-m, -\epsilon)$, being $\epsilon= B_z/B_\phi$ and $m$ the azimuthal wavenumber.
This has led to the speculation that this instability can produce a {\it magnetic} $\alpha$-effect which, in turn, could produce a dynamo action \citep{spruit02,ruki12}. 
The effect of a finite Prandtl number has been investigated in cylindrical geometry \citep{ruge12}, while the role played by a finite electrical resistivity in liquid conductor has been discussed in \cite{weber13}. 
A recent work in this direction \citep{weber15} has discussed the numerical evidence that the saturation state of the Tayler instability at low magnetic Prantdl number and high Hartmann number is characterized by helicity oscillations; a result of important consequences if confirmed \citep{stefani15}. 

The aim of this paper is to extend the approach to the  non-linear saturation phase of the Tayler instability discussed in \cite{bonanno12} in order to account for additional non-linearities possibly induced by finite Prandtl number in the system discussed by \cite{weber15}. 
In particular, if the selection of a final helical state occurs via the mechanism of spontaneous symmetry breaking and mutual antagonism (a well-known phenomenon occurring in various contexts \citep{dels12}) one can argue that the zoo of possible final states in more complicated situations can also be classified in terms of fixed points of the dynamical system for the amplitude equations.
This approach is very general and can be easily extended to include further non-linearities and additional couplings due to explicit symmetry breaking terms. Our main goal is to study the robustness of the findings of \citep{bonanno12} by including higher-order terms in the effective Lagrangian for the weakly non-linear amplitude 
coupling. In particular, we would like to see if the findings of \citep{weber15} can show up in the phase space 
as a limit-cycle characterized by an oscillating (in time) helical state. 

The plan of the paper is the following: in Sec. 2 we introduce the basic formalism, in Sec. 3 we discuss the case
of spontaneous symmetry breaking, in Sec. 4 the possible modification of the phase space due to an explicity symmetry
breaking, and Sec. 5 is devoted to the conclusions. 

\section{Basic formalism} 
\begin{figure}[ht]
\begin{center}
\includegraphics[width=5cm]{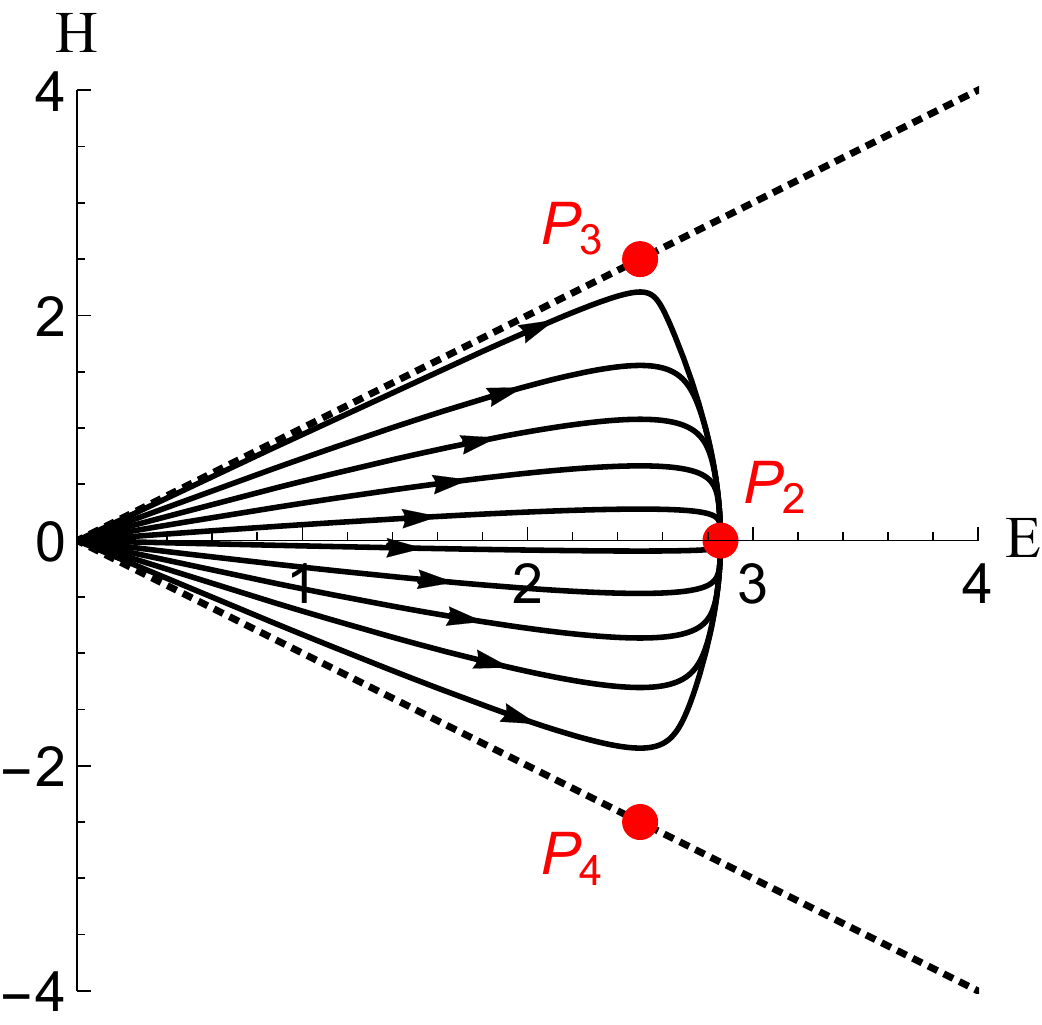}
\includegraphics[width=5cm]{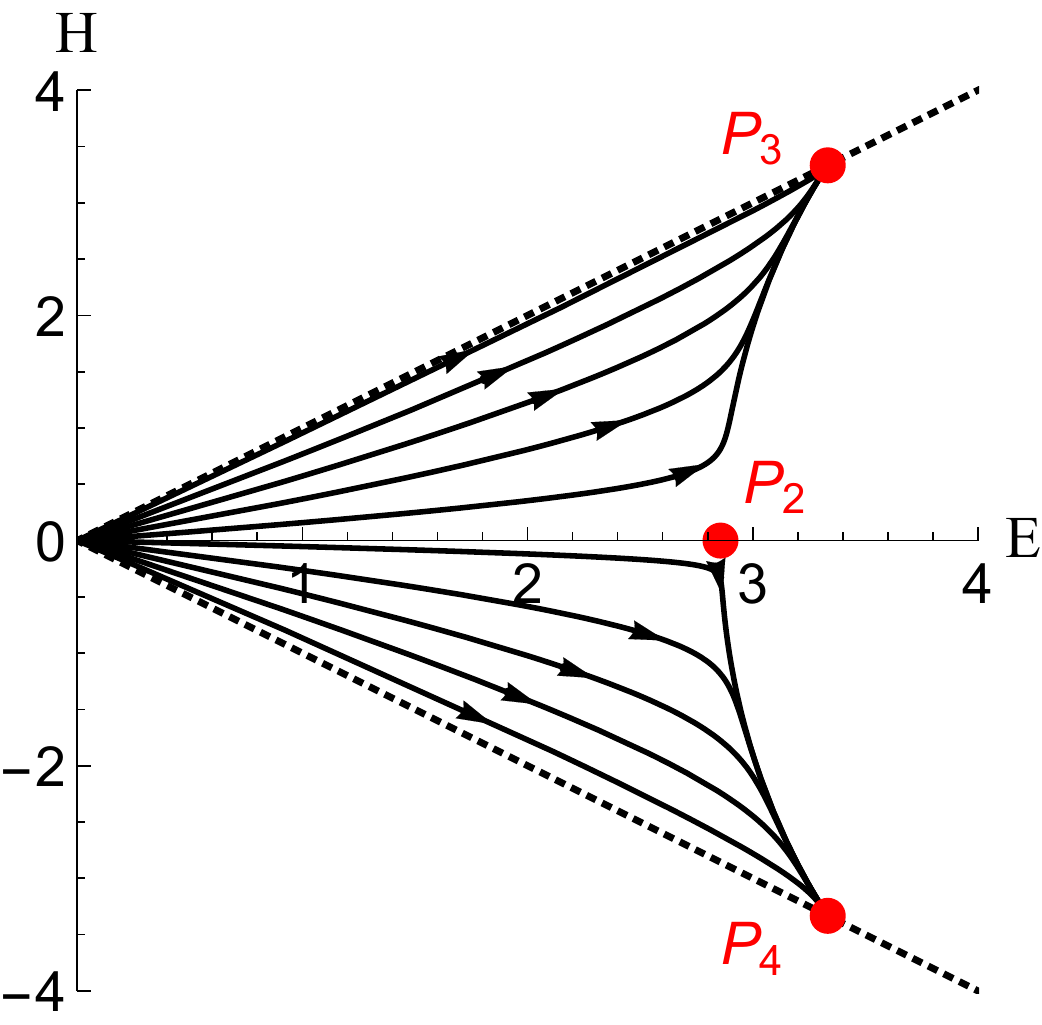}
\caption{Upper panel: $\mu^* = 0.15, \mu = 0.2, \gamma =1$. Lower panel: $\mu^* = 0.2, \mu = 0.1, \gamma =1$.}
\label{fig:plot4}
\end{center}
\end{figure}

Let us consider an ideal MHD system in cylindrical geometry with cylindrical coordinates $\bx = \{r,\phi,z\}$ and axial symmetry $B(\bx) \equiv B(r)$, and in particular a static ground state characterized by a magnetic field $B_0(r) = (0, B_\phi(r),B_z(r))$ balanced by a fluid pressure so that
\be
\frac{\nabla\, p_0}{\rho} = F_L = \frac{1}{4\, \pi_0\, \rho}\, (\nabla \times B_0) \times B_0\, .
\ee
We want to study the evolution of Tayler instability after perturbing the ground state with a non-axisymmetric perturbation\footnote{Note that the perturbation consists of two fields $B_1$ and $v_1$, respectively the perturbations for the magnetic and velocity fields. We refer to $B_1$ as the perturbation since in the linearized equations the two fields are linearly dependent \citep{bour08}.} $B_1$. As it is well known, the linearized system can be investigated in spatial Fourier space by employing a dependence from $t$, $\phi$ and $z$ of the type $exp(\gamma\, t - i\, \phi\, m - i\, k_z\, z)$, being $m$ and $k_z$ respectively the azimuthal wavenumber and the wave vector in the axial direction. We then end up with an eigenvalue problem whose solution is a spectrum of infinite unstable eigenmodes, 
with same growth rate $\gamma$ for pairs of opposite $m$, and with the most unstable eigenfunction for $m=\pm1$ \citep{bonanno11,bonanno12}. While any axisymmetric perturbation ($m=0$) preserves the symmetries of the ground state, an helical perturbation may lead to a state with finite helicity. The latter is an interesting example of spontaneous symmetry breaking, i.e. when the ground state does not satisfies the symmetries of the system. The eigenfunction are found to be linear superpositions of Bessel functions \citep{tataronis87}, that have a long history in plasma physics. \cite{lund} has shown that force-free  solutions satisfy the Beltrami relation
\be\label{hel1}
\nabla \times \bB(r) = \alpha(r)\, \bB(r) \, , 
\ee
and in particular the solution stays force-free along its time-evolution if it is helical, i.e. $\alpha(r) = const$, and read $B_\phi = A\, J_1(\alpha\,r), B_z = A\, J_0(\alpha\, r)$, being $J$ the Bessel function of first kind and $A$ a free parameter. Chandrasekhar and Woltjer
\citep{chandrasekhar1958} also proved that these solutions maximize the helicity for a given energy. Helical solutions have been seen in recent simulations at high Prandtl numbers $Pm = 10^{7}$ by \cite{bonanno11}, and at $Pm = 0.1, 1, 10$ by \cite{gellert11}. 
Simulations did show, however, that helicity starts to grow but ultimately goes to zero after a transient, hence not reaching saturation. 
In particular, in \cite{rudiger11a,rudiger11b} it has been shown that for a constant current density 
in an infinitely long cylinder the governing parameter is the Hartmann number, $Ha = B_\phi(r_{out})\,r_{out}(\sigma/\rho \nu)^{1/2}$, with $\sigma$, $\rho$, and $\nu$ respectively the conductivity, density and viscosity of the fluid, instead of Lundquist number $S = Ha\, Pm^{1/2}$. 
Simulations at high Hartmann numbers \citep{weber15,stefani15} have shown production of helicity in Tayler instability that does not always decay to zero but may also saturate to a finite value presenting in both cases new features like oscillations and limit cycles. While simulations are discovering an apparently broad family of saturation patterns, not much can be said from the analysical point of view.
While in the linear regime the perturbation depends solely on the initial helicity and growth rate, 
the final state is generated by the large eigenfunction mixing due to the non-linearities of the system. 
A qualitative picture of the final states may be however obtainable by exploiting solely the symmetries of the system. \cite{bonanno11} reproduced the spontaneous symmetry breaking by employing a simple model based on symmetries. 
The model is based on a effective Lagrangian invariant under a helicity-parity transformation $\mathcal{P}_{LR}$: $m \to -m$, which is achieved in the limit of vanishing $B_z$-component, i.e. $\Lim{B_z\to0} P_{LR}{}^z = P_{LR}{}$. The chiral simmetry governs the equation of motion of two modes with oppositive $m$, and hence opposite helicity and same growth rate.  
In Fourier space these modes read
\ba
&&L(\bx) = \sum \bL(\bk)\, \psi(\bk, \bx)\, d\bx\\
&&R(\bx) = \sum \bR(\bk)\, \psi(\bk, \bx)\, d\bx ,
\ea
where the Bessel-Fourier basis $\psi$ reads
\ba\label{Chabase}
\psi(\bx,\bk) = e^{\,i\, m\, \phi + i\, k_z\, z}  J_m\left(r\, \sqrt{k^2 - n_z^2}\right)  \, ,
\ea
being $J$ the Bessel function of the first kind, $m=\pm1,2,3,\cdots$, and where $k$ is the eigenvalue solution of the Helmholtz equation $\nabla^2\, \psi(\bx) = k^2\, \psi(\bx)$, $\nabla^2$ being the Laplacian. In the limit $r_{in}\to0$, $r_{out}\to\infty$, respectively the inner and outer radiuses of our cylindrical geometry, (\ref{Chabase}) is a complete set of orthonormal functions, i.e.
\be\label{orthon}
\int_0^{\infty}\, dr\, J_{m}(k_1\, r)\, J_{m'}(k_2\,r) = \frac{1}{k_1}\, \delta_{m\, m'}\,\delta(k_1 - k_2) \, .
\ee
As the basis is complete, any initial condition can be parametrized in terms of linear combination of basis elements. Eq.(\ref{orthon}) also implies orthonormality between left and and right modes (i.e. $m$ and $-m)$. Being helical, the modes satisfy
\ba\label{hel2}
&&\nabla \times L(\bx) = \alpha\, L(\bx) \, , \\
&&\nabla \times R(\bx) = - \alpha\, R(\bx) \, . 
\ea
Left and right energy and helicity densities can then be defined as
\ba
&&E_L = \frac{1}{2}\, \int L(x)^2\, d\bx = \frac{1}{2}\,\bL\, \bL^* \,,\\
&&E_R = \frac{1}{2}\, \int R(x)^2\, d\bx = \frac{1}{2}\,\bR\, \bR^* \,,\\
&&\mathcal{H}_L = (\nabla \times L) \cdot L \, ,\\
&&\mathcal{H}_R = (\nabla \times R) \cdot R\, , 
\ea
where the asterisk denotes complex conjugation, so that using eq.(\ref{hel1}) we obtain
\be\label{helen}
\mathcal{H}_L = 2\, \alpha\, E_L\, , \qquad \mathcal{H}_R = - 2\, \alpha\, E_R\, .
\ee
The effective Lagrangian in the weakly non-linear case reads
\ba\label{lagrangian}
\mathcal{L}(\bL,\bR) = \\\frac{1}{2}\, \gamma\, (\bL^2 + \bR^2) - \frac{1}{4}\, \mu\, (\bL^4 + \bR^4) - \frac{1}{2}\, \mu^*\, (\bL^2\, \bR^2) + \mathcal{O}(4)\, \nonumber,
\ea
where $\mathcal{O}(4)$ includes higher-order terms in $\bL^4$ and $\bR^4$. Eq.(\ref{helen}) is the most general fourth-order Ginzburg-Landau type of Lagrangian compatible with the helical symmetry $\mathcal{P}_{LR}$.
The evolution equations for the two modes read then
\ba\label{motion}
&&d_t \, \bL = \frac{\delta\, \mathcal{L}(\bL,\bR)}{\delta \bL}\,, \\
&&d_t \, \bR = \frac{\delta\, \mathcal{L}(\bL,\bR)}{\delta \bR}\, .
\ea
The above equations lead to a couple of equations for the left and right energy densities introduced in (\ref{helen}), i.e.
\ba
&&d_t\, \text{E}_L = 2\, \gamma\, \text{E}_L - 4\, \mu\ \, \text{E}_L^2 - 4\, \mu^*\, \text{E}_L\, \text{E}_R\, , \nonumber \\
&&d_t\, \text{E}_R = 2\, \gamma\, \text{E}_R - 4\, \mu\ \, \text{E}_R^2 - 4\, \mu^*\, \text{E}_L\, \text{E}_R\, ,
\ea
and therefore analogous equations for the left and right helicities.
We then define the total helicity and energy respectively as $\mathcal{H} = \mathcal{H}_L + \mathcal{H}_R$ and $E = E_L + E_R$, introduce the rescaled total helicity $H = \mathcal{H}/2\,\alpha$, so that the left and right energy densities read $E_L = E - H$ and $E_R = E + H$. 
At last we have
\ba\label{EHsys4}
&&d_t\, \text{H} = 2\, \gamma \text{H}\, - 4\, \text{H}\, \text{E}\, \mu\,  \nonumber ,  \\ 
&&d_t\, \text{E} = 2\, \gamma\, \text{E} - 2\, (\mu + \mu^*)\, \text{E}^2 - 2\, (\mu - \mu^*)\, \text{H}^2 \, .
\ea
The system (\ref{EHsys4}) has four fixed points. 
A trivial solution at $P_1 = \{E^*{}_1 = 0,  H^*{}_1 = 0\}$, a point $P_2 = \{E^*{}_2 = 2\, \gamma\, ( \mu + \mu^*), H^*{}_2 = 0\}$, and a pair of helicity-maximizing points $P_3 = \{E^*{}_3 = E_s, \,  H^*{}_3 = E_s\}, P_4 = \{E^*{}_3 = E_s, \,  H^*{}_3 = - E_s\}$, 
with $E_s = \gamma/2\, \mu$. Their eigenvalues are respectively $\Theta^{(1)}{}_1 = \Theta^{(1)}{}_2 = 2\, \gamma$, $\Theta^{(2)}{}_1 = - 2\, \gamma, \Theta^{(2)}{}_2 = 2\, \gamma - 4\, \gamma\, \mu/(\mu + \mu^*)$, $\Theta^{(3)}{}_1 = \Theta^{(4)}{}_1 = - 2\, \gamma, \Theta^{(3)}{}_2 = \Theta^{(4)}{}_2 = 2\, (\mu - \mu^*)/\gamma$. 
The origin is always repulsive while the point $P_2$ and the couple $P_3$ and $P_4$ are sink or saddle point depending on the values of $\mu$ and $\mu^*$. 
As depicted in Fig.\ref{fig:plot4}, the physical parameter space is delineated by the cone $H<E$ with $E>0$. For $\mu <\mu^*$ the point $P_2$ is attractive, representing an achiral final state (symmetric phase), while for $\mu>\mu^*$ the system reaches a helical configuration represented by either the point $P_3$ or $P_4$ (broken phase). 
The selection of a helical states in our model closely resembles the chiral symmetry breaking occurring in biochemistry with mirror-simmetric  
bio-molecules (mainly sugars and amino acids) \citep{saito13}. 

\section{Spontaneuos Symmetry Breaking} 
At the next order there are two additional terms which are invariants under $L\to R$ parity, respectively 
$\lambda\, (L^6+R^6)$ and $\lambda^*\,(L^2\, R^4 + R^2\, L^4)$. The Lagrangian thus reads 
\ba\label{lagrangian2}
\mathcal{L}(\bL,\bR)& = &\frac{1}{2}\, \gamma\, (\bL^2 + \bR^2) - \frac{1}{4}\, \mu\, (\bL^4 + \bR^4) 
- \frac{1}{2}\,\mu^*\, (\bL^2\, \bR^2)\nonumber \\  
&-& \frac{1}{6}\, \lambda\, (\bL^6 + \bR^6) - \lambda^*\, (\bL^2\, \bR^4 + \bL^4\, \bR^2)\, .
\ea
The time evolution of the left and right mode in this case is ruled by 
\ba\label{LR6}
&&d_t\, \bL = \gamma\, \bL -  \lambda\, \bL^5 - \mu\, \bL^3 - 
\lambda^*\, (4\, \bL^3\, \bR^2 + 2\, \bL\, \bR^4 ) - \mu^*\, \bL\, \bR^2 \nonumber \, ,\\
&&d_t\, \bR =   \gamma\,  \bR - \lambda\, \bR^5 - \mu\, \bR^3 - 
\lambda^*\, (4\, \bL^2\, \bR^3 + 2\, \bL^4\, \bR ) - \mu^*\, \bL^2\, \bR \nonumber \, ,\\
\ea
which leads to the following dynamical system for energy and helicity densities
%
\ba\label{sys6}
&&d_t\, \text{H} = - 2\, \{- \text{H}\, \gamma + \text{H}\, \text{E}^2\, \left(2\, \lambda^* + 3\, \lambda\right)  \nonumber \\
&&\hspace{2em} + 2\, \text{H}\, \text{E}\, \mu + \text{H}^3\, (\lambda - 2\,\lambda^*)\} \,,\\
&&d_t\, \text{E} = - 2\, \{\text{E}^3\, \left(6\, \lambda^* + \lambda \right) + \text{E}^2\, \left(\mu^* + \mu \right) \nonumber \\
&&\hspace{2em}- \text{E}(\gamma - 3\, \text{H}^2\, (\lambda - 2\, \lambda^*)) + \text{H}^2\, (\mu - \mu^*) \}\, .
\label{sys6d}
\ea
%
Eqs.(\ref{sys6}-\ref{sys6d}) have nine fixed points, $P_i = \{E^*{}_i, H^*{}_i\}$, where $i=\{1,\cdots, 9\}$. Of these\footnote{The analytic expressions of the fixed points and their eigenvalues can be found in the appendix A.}, one is the trivial solution $P_1 = \{0, 0\}$, with eigenvalues unchanged from the weakly nonlinear case, $\Theta^{(1)}{}_1 = \Theta^{(1)}{}_2 = 2\, \gamma$; two fixed points lie on the $H=0$ line, respectively $P_2 = \{E^*{}_2, 0\}$ and $P_3 = \{E^*{}_3, 0\}$; 
two points, $P_4$ and $P_5$, 
are helicity-maximizing solutions ($E^* = \pm \,H^*$); the remaining four points $P_6,\cdots, P_9$, correspond to saturated configurations where $E^* \neq H^*$. 
The points $P_1, P_3, P_4$ and $P_5$ are higher-order generalizations of the weakly nonlinear case, as it can be seen by taking the double limit $\lambda \to \lambda^* \to 0$ of eq.\ref{ssbfp}. Of the remaining points not all are physical. The points $P_2$, $P_6$ and $P_7$ lie always in the negative energy region, while the remaining two points, $P_8$ and $P_9$ satisfy the constraint $H^*<E^*$ only in a narrow window of the parameter space. 
The presence (or absence) of those two new fixed points leads to two qualitatively different phase portraits.
\begin{figure}[ht]
\begin{center}
\includegraphics[width=5cm]{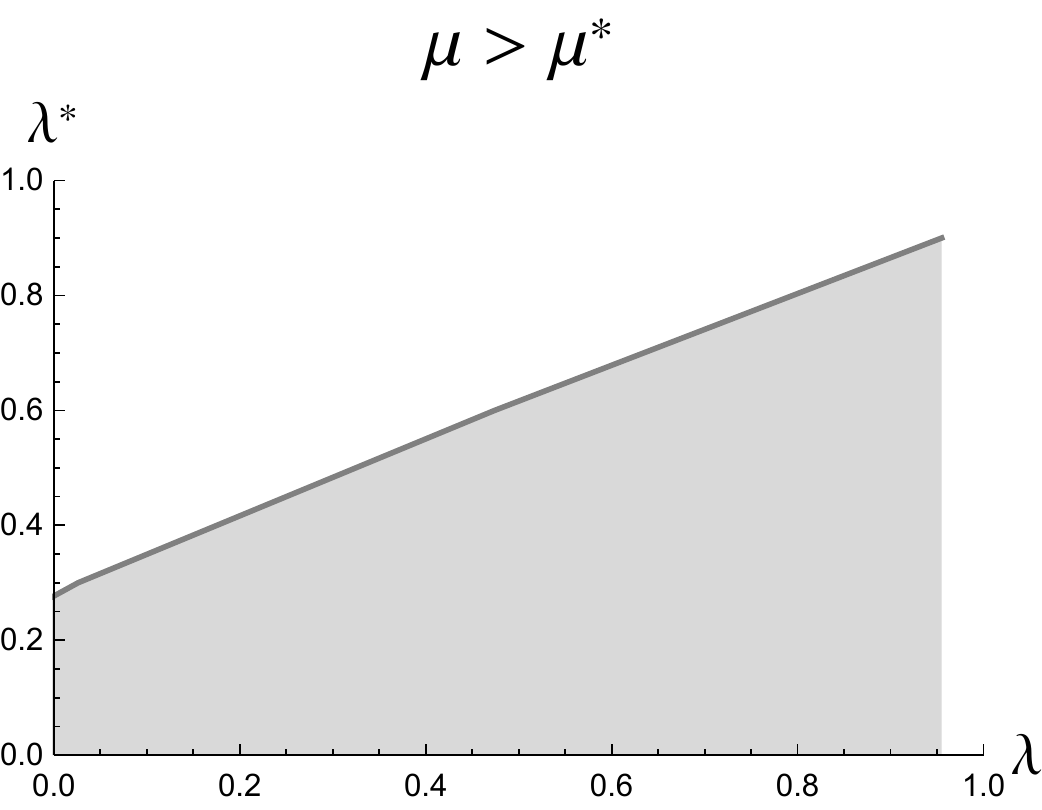}
\includegraphics[width=5cm]{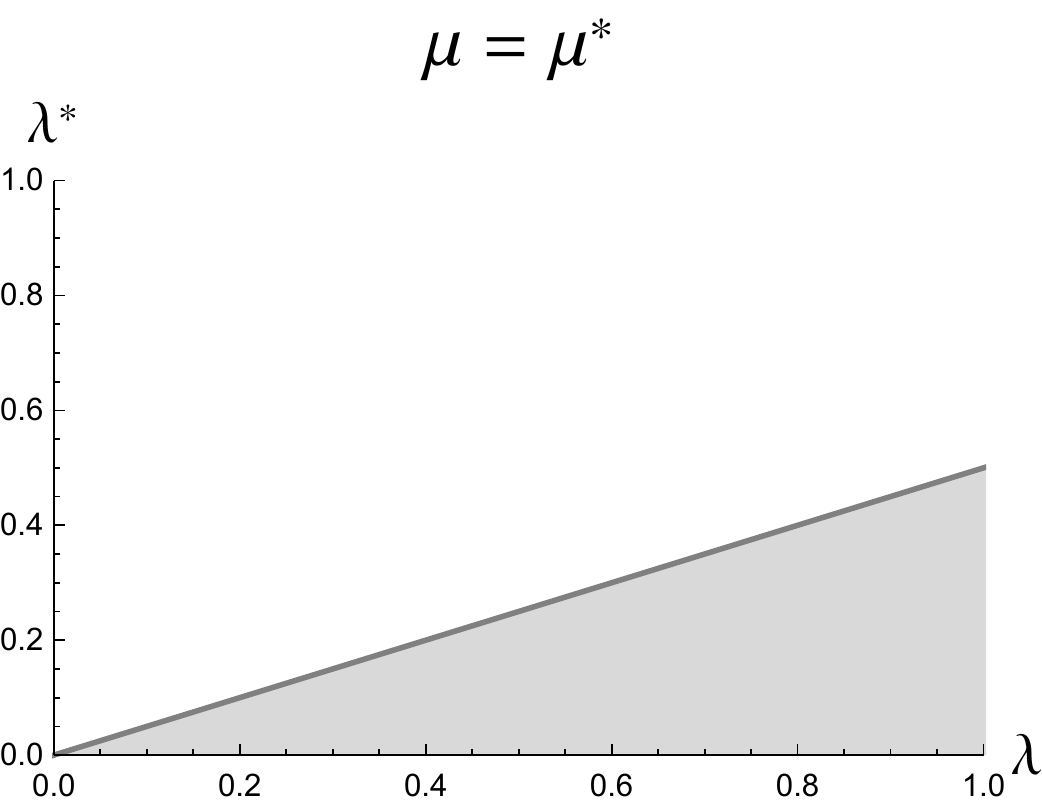}
\includegraphics[width=5cm]{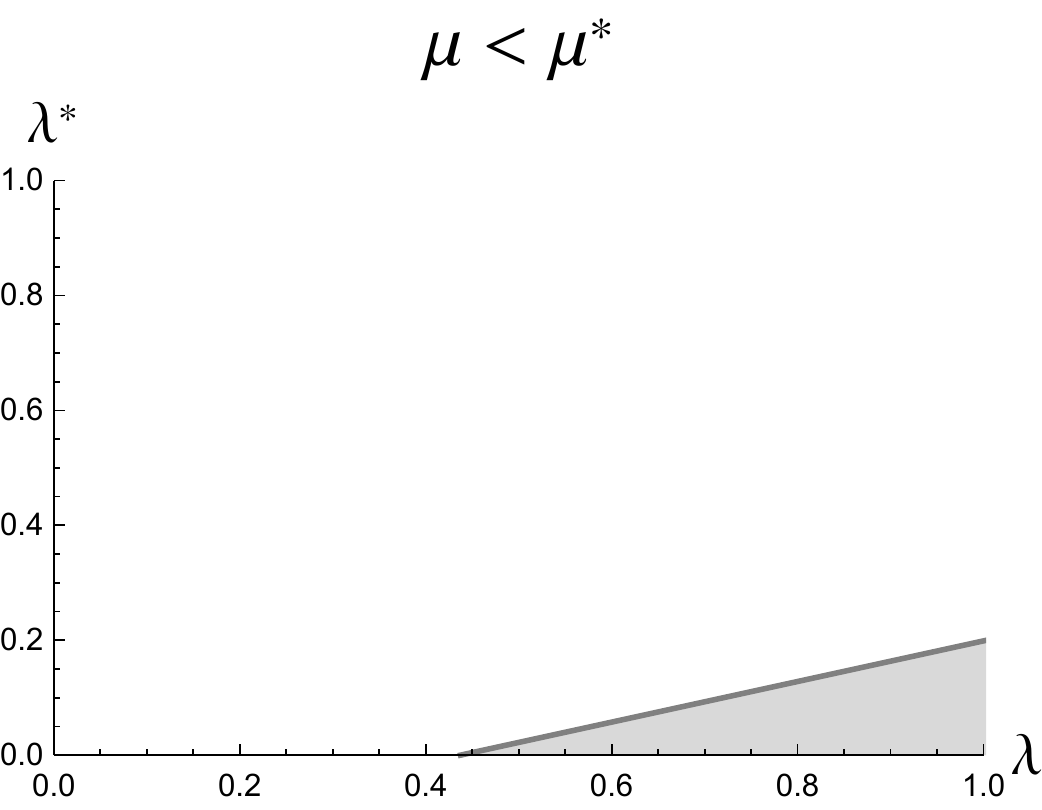}
\caption{From top to bottom: $\{\mu=1, \mu^*=0.5\}, \{\mu=1, \mu^*=1\}, \{\mu=0.5, \mu^*=1\}$. 
Symmetric and broken phase are realized respectively in the gray and white zone.}
\label{fig:plotstability}
\end{center}
\end{figure}
In their absence, the phenomenology is similar to that of the weakly interacting case, that has two phases, a symmetric one at $\mu > \mu^*$ ($P_3$ is a sink, $P_4$ and $P_5$ are saddle points) and a broken phase at $\mu < \mu^*$ ($P_3$ is a saddle point, $P_4$ and $P_5$ are sinks). 
In the higher-order case, the critical line $\mu - \mu* = f(\mu,\mu^*) = 0$ generalizes to a critical hypersurface $f(\mu, \mu^*,\lambda,\lambda^*)=0$ that is function of the new parameters\footnote{Note that, because of the larger parameter space, $\mu$ and $\mu^*$ can now assume negative values. We did not notice qualitative differences between the positive and negative regions of parameter space.}. 
The critical surface is depicted in Fig.\ref{fig:plotstability} for fixed values of $\mu$ and $\mu^*$.

\begin{figure}
\begin{center}
\includegraphics[width=4cm]{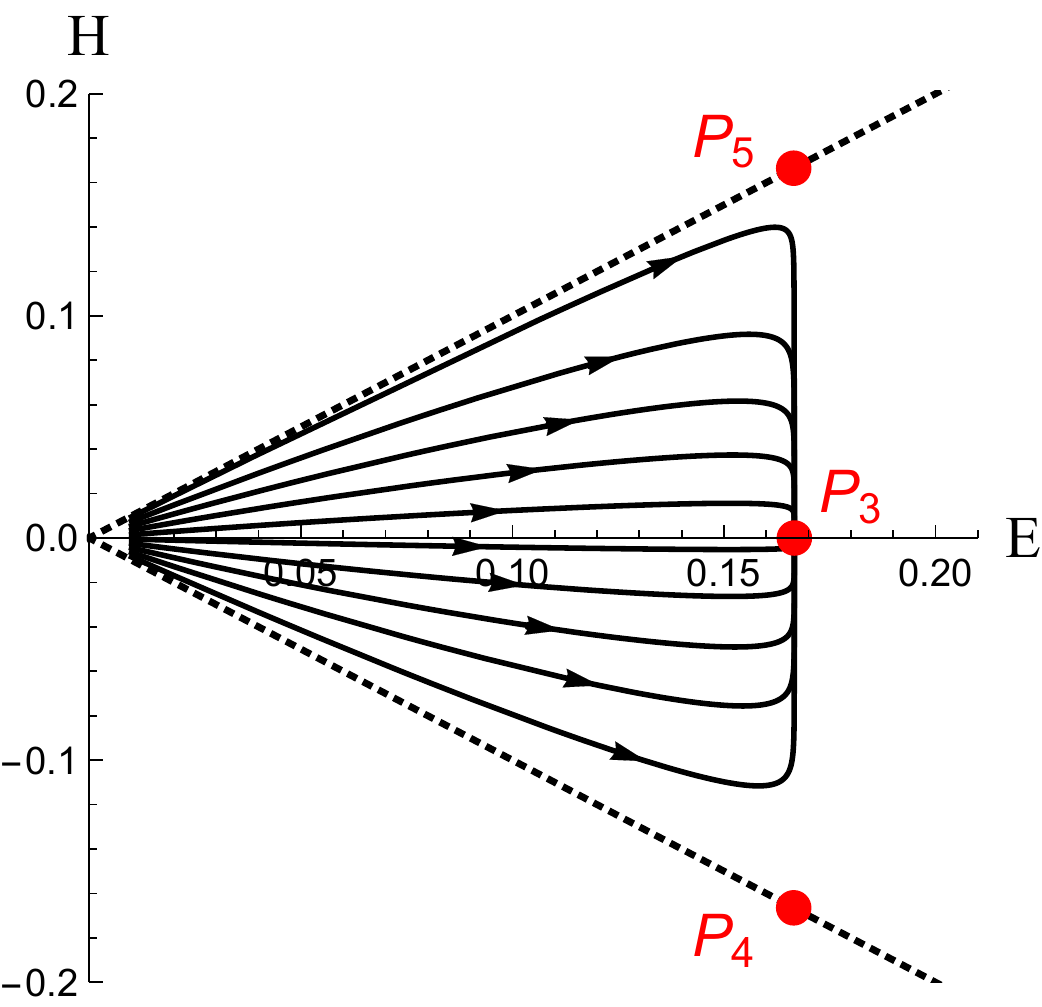}  
\includegraphics[width=4cm]{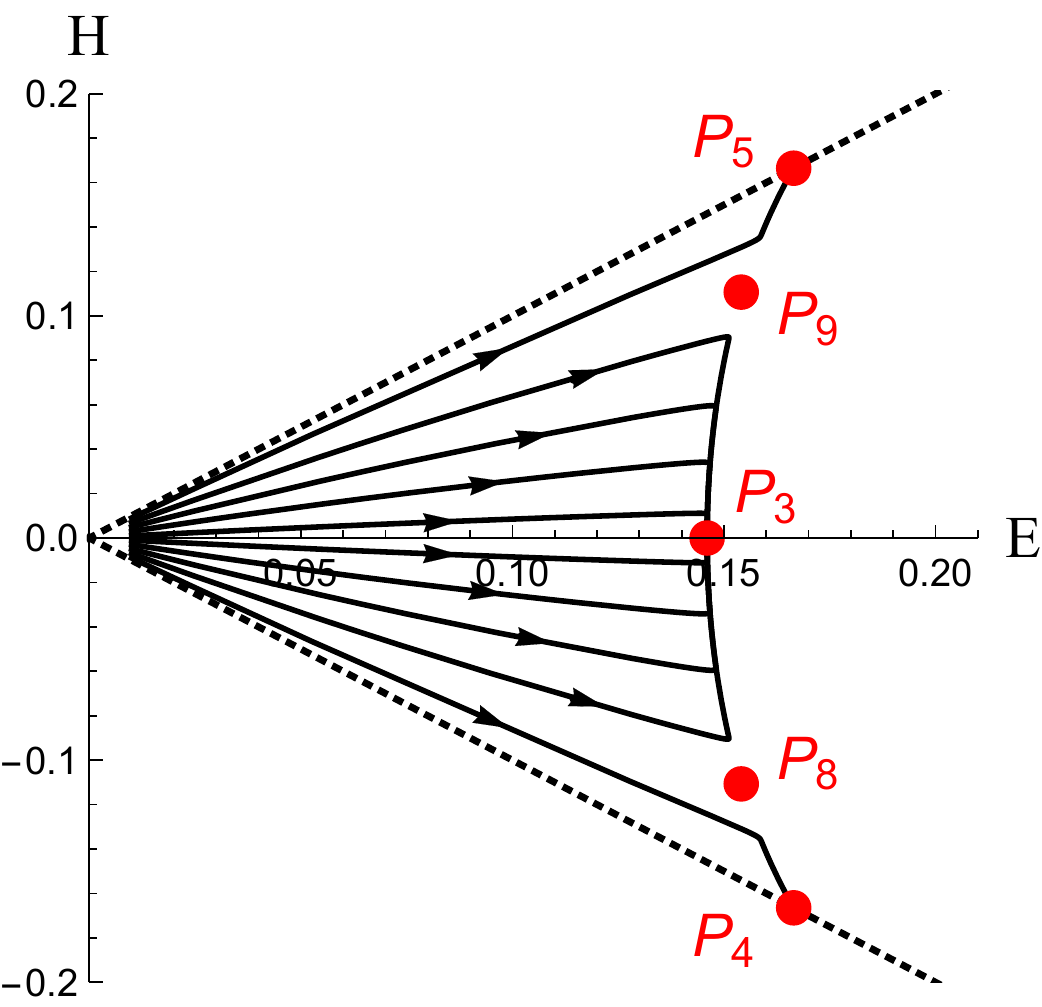}
\includegraphics[width=4cm]{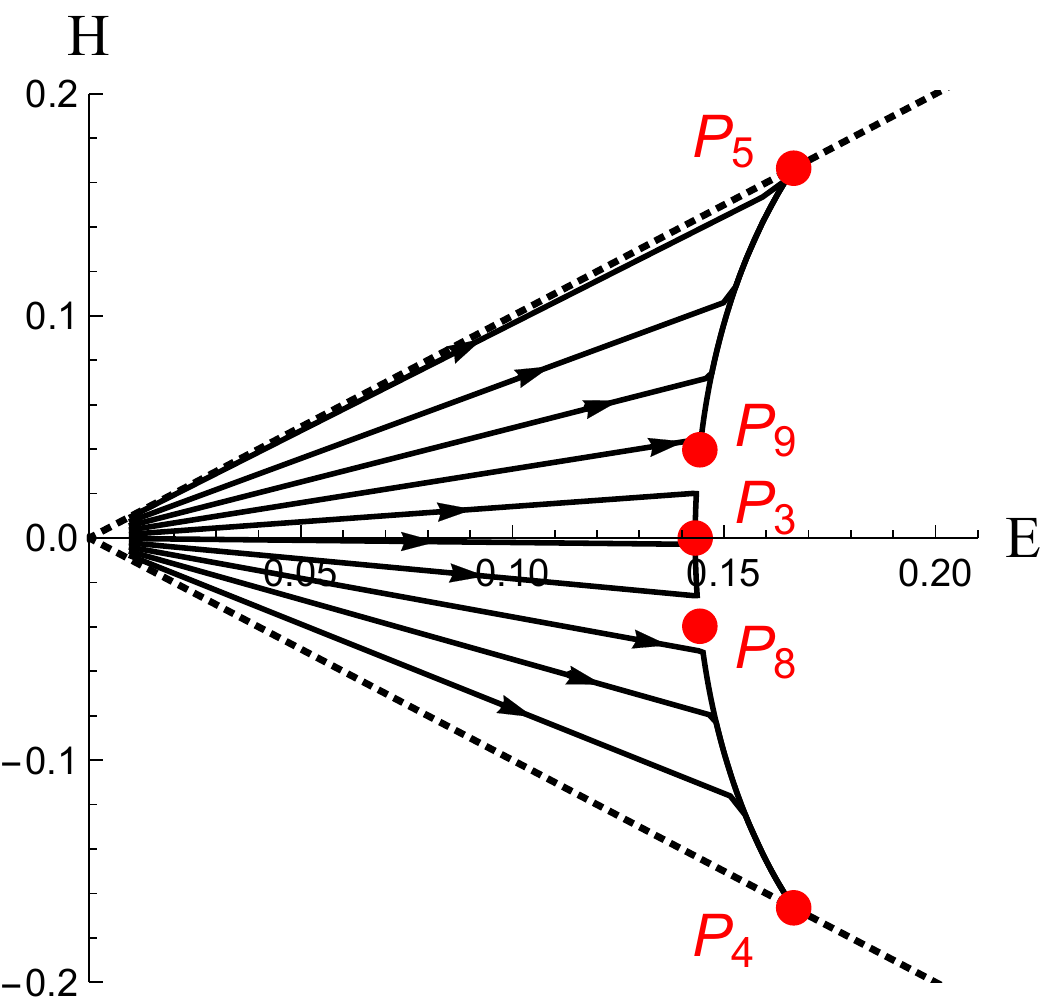}  
\includegraphics[width=4cm]{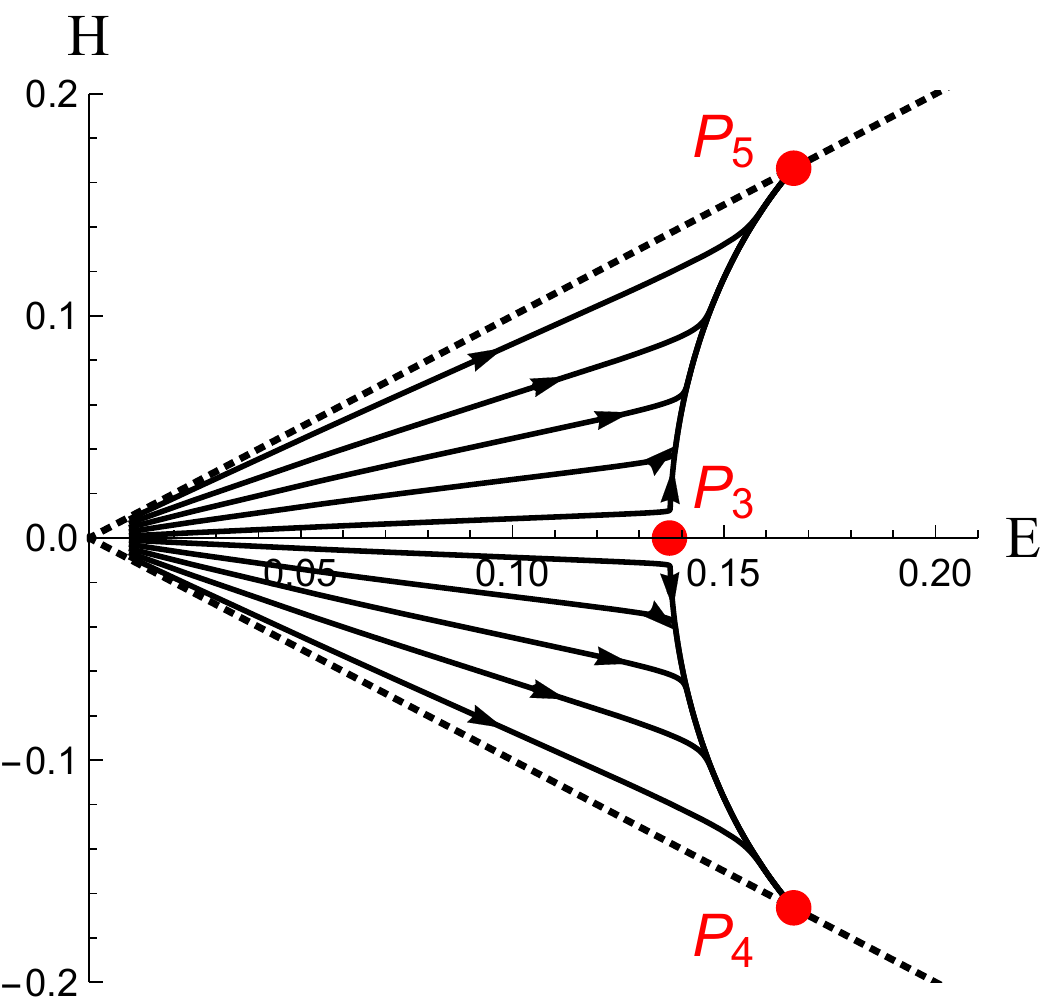}
\end{center}
\caption{From top-left to bottom-right: $\lambda^* = 2, 3.3, 3.47, 4.0.$ The other parameters are set at $\mu= 3, \mu^* = 1, \gamma = 1, \lambda = 0.01$.}
\label{fig:plotcone}
\end{figure}

\begin{figure}
\begin{center}
\includegraphics[width=4cm]{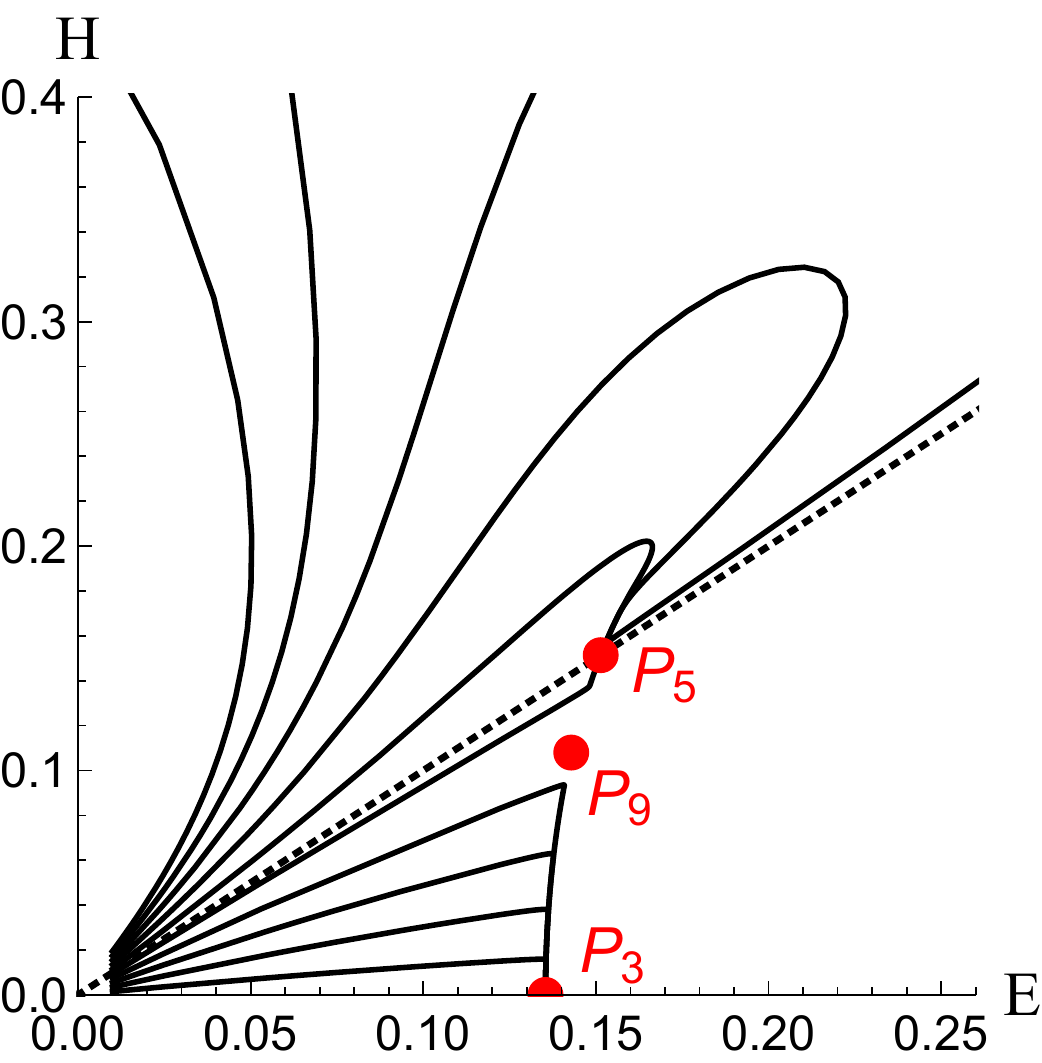}
\includegraphics[width=4cm]{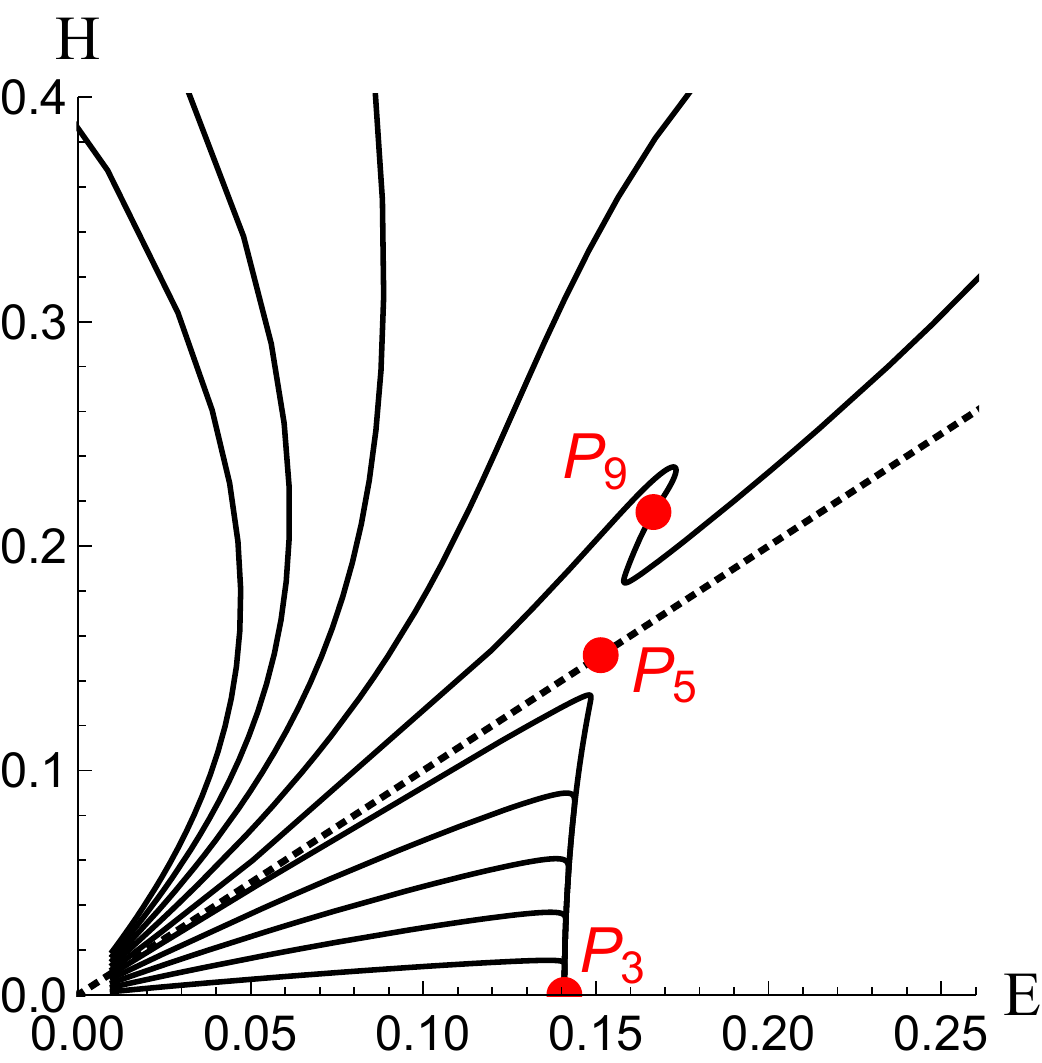}
\includegraphics[width=4cm]{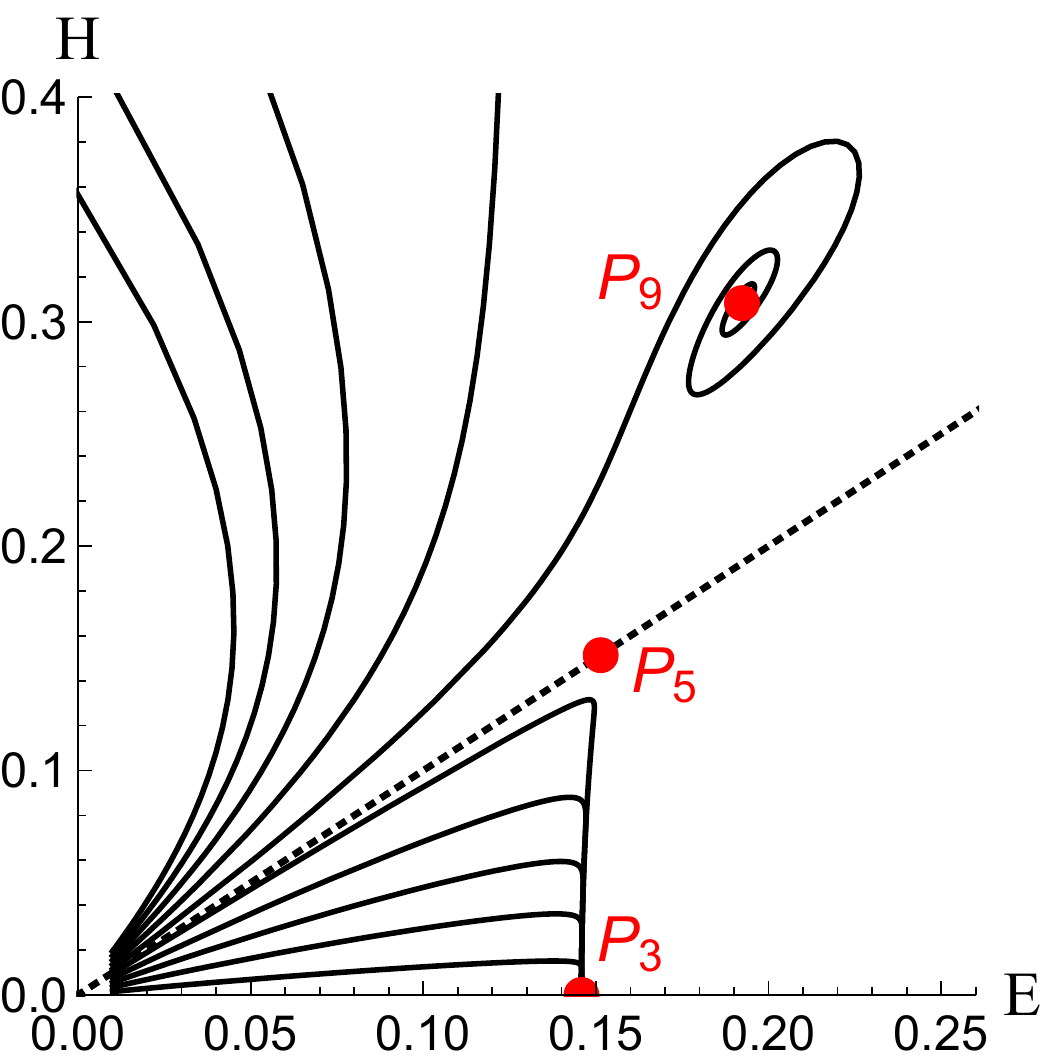}
\caption{From top-left to bottom-right: $\lambda^*=\{4, 3.5, 3.1\}$. The other parameters are set to $\lambda=1, \mu^*=1, \mu=3$.}
\label{fig:plotspiral}
\end{center}
\end{figure}

Let us consider the symmetric phase in the weakly interacting case, 
i.e. $\mu > \mu^*$ and $\lambda = \lambda^* = 0$. 
Turning on the higher-order parameters, the symmetric phase (gray zone in Fig.\ref{fig:plotstability}) endures for 
\be\label{lcrit1}
\lambda > \lambda_{cr}(\mu,\mu^*,\lambda^*)\, ,
\ee
that is generally satisfied for $\lambda \gg \lambda^*$, and where $\lambda_{cr}$ can be obtained by solving $f(\mu, \mu^*,\lambda ,\lambda^*)=0$ respect to $\lambda$, and is solution either of $\Theta^{(4)}{}_1 = \Theta^{(5)}{}_1 = 0$ or $\Theta^{(3)}{}_2 = 0$. 
In the broken phase, where $\mu < \mu^*$, 
the helical solutions are stable (white zone in Fig.\ref{fig:plotstability}) for
\be\label{lcrit2}
\lambda < \lambda_{cr}(\mu,\mu^*,\lambda^*)\, ,
\ee
realized for $\lambda \ll \lambda^*$. In all but a small windows of the parameter space $\{\mu,\mu^*,\lambda,\lambda^*\}$, the function $\lambda_{cr}$ 
evaluated by either the eigenvalues of the sink or the saddle point coincide, hence indicating a sharp phase transition. In a small region in parameter space, however, the two solution do not overlap, and instead $\lambda^{(3)}_{cr} < \lambda^{(4,5)}_{cr}$. 
In this region $P_8$ and $P_9$ are physical saddle points. 
As depicted in Fig.\ref{fig:plotcone}, for $\lambda < \lambda^{(3)}_{cr}$ $P_8$ and $P_9$ are unphysical and the helical fixed points are sinks (bottom-right panel). 
In the region $\lambda^{(3)}_{cr} <\lambda < \lambda^{(4,5)}_{cr}$ 
the helical and non-helical fixed points are both sinks, while $P_8$ and $P_9$ are saddle points (bottom-left and top-right panels), for $\lambda > \lambda^{(4,5)}_{cr}$ the points $P_8$ and $P_9$ become complex and the non-helical point is the only sink (top-left panel).
The latter phase portrait opens to an interesting scenario in which 
both the fully-helical and non-helical fixed points are of saddle, 
and $P_8$ and $P_9$ are the only sinks. The eigenvalues of these fixed points, moreover, may a priori be a complex conjugated pair, hence showing non-trivial behaviors like oscillations and limit cycles. 
However, an extensive analysis in parameter space showed no presence of the case here suggested, which is instead realized, together with complex eigenvalues, outside of the physical regime, how depicted in Fig.\ref{fig:plotspiral}. 
Interestingly, a similar situation (which includes oscillations in the unphysical regime) can be instead realized starting from a non-axisymmetric ground state. 
%

\section{Explicit Symmetry Breaking} 
\begin{center}
\begin{figure}[ht]
\includegraphics[width=5cm]{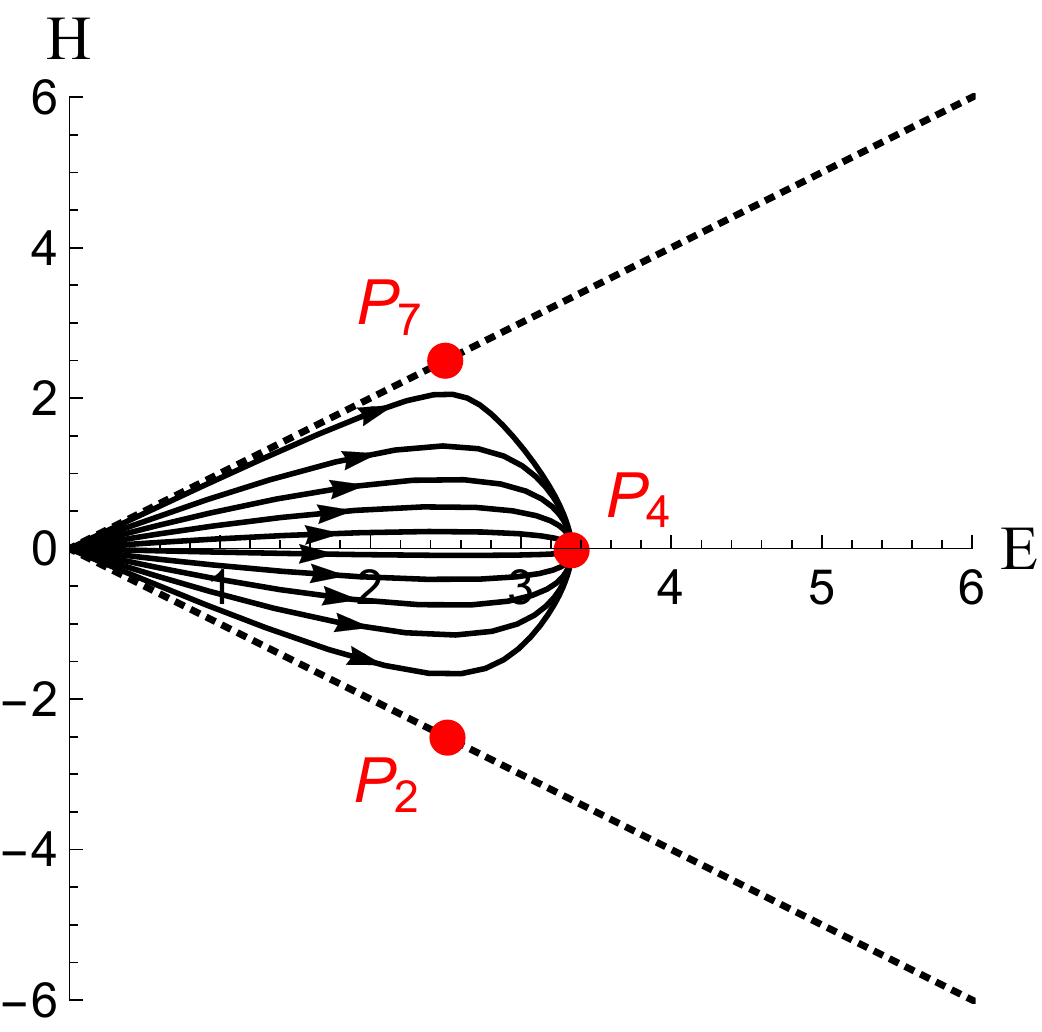}
\includegraphics[width=5cm]{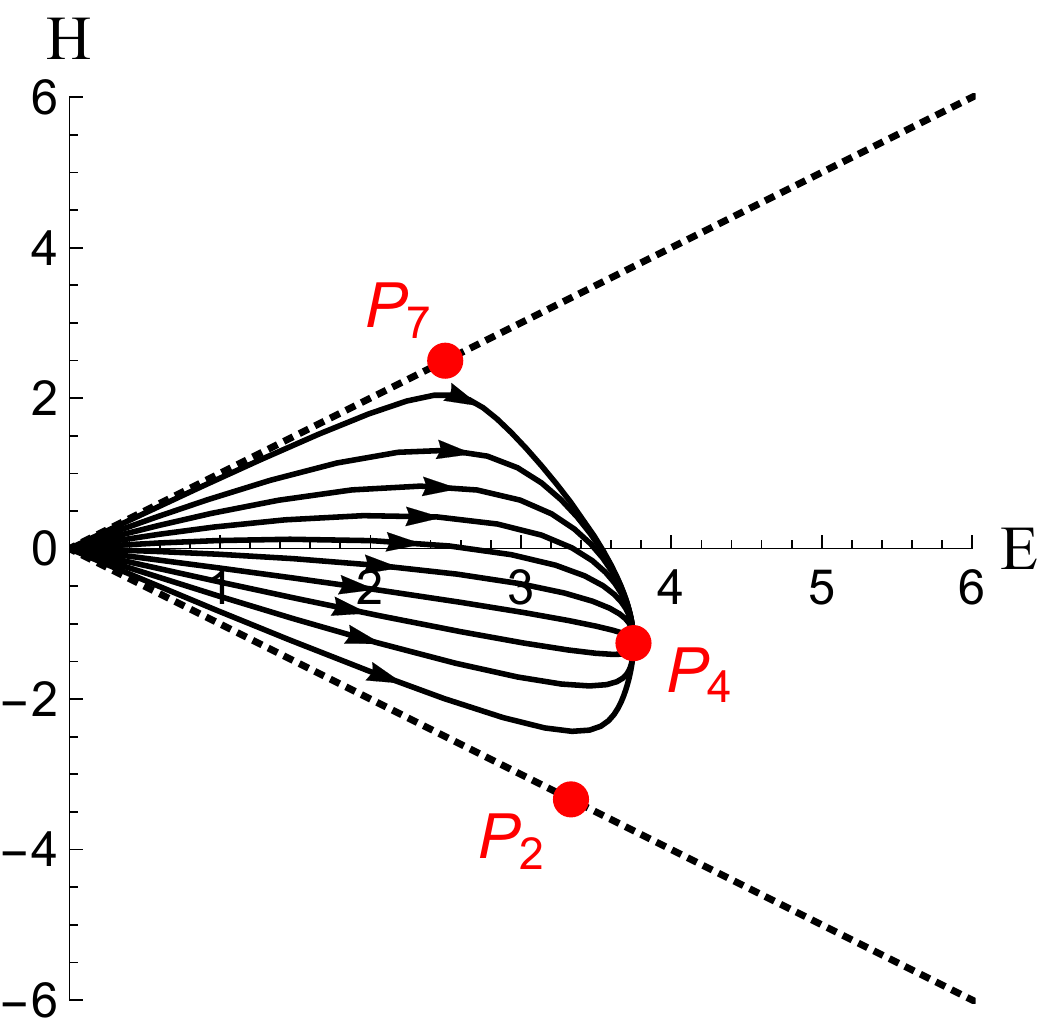}
\includegraphics[width=5cm]{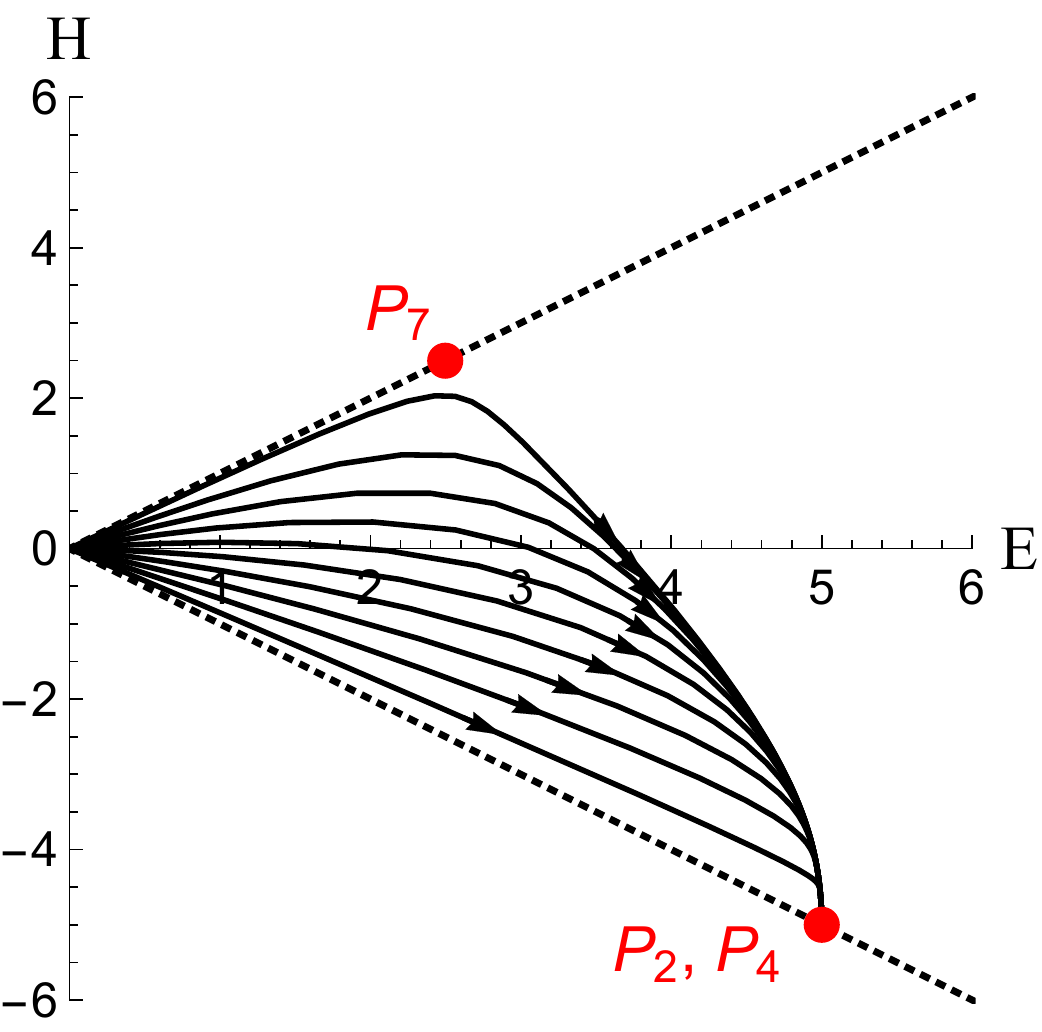}
\caption{From top to bottom: $\mu_L=0.20, 0.15, 0.10.$ The other parameters are set at $\mu_R = 0.2, \gamma_L = \gamma_R= 1, \mu^* = 0.1$.}
\label{fig:plotordered}
\end{figure}
\end{center}

\begin{center}
\begin{figure}[ht]
\includegraphics[width=5cm]{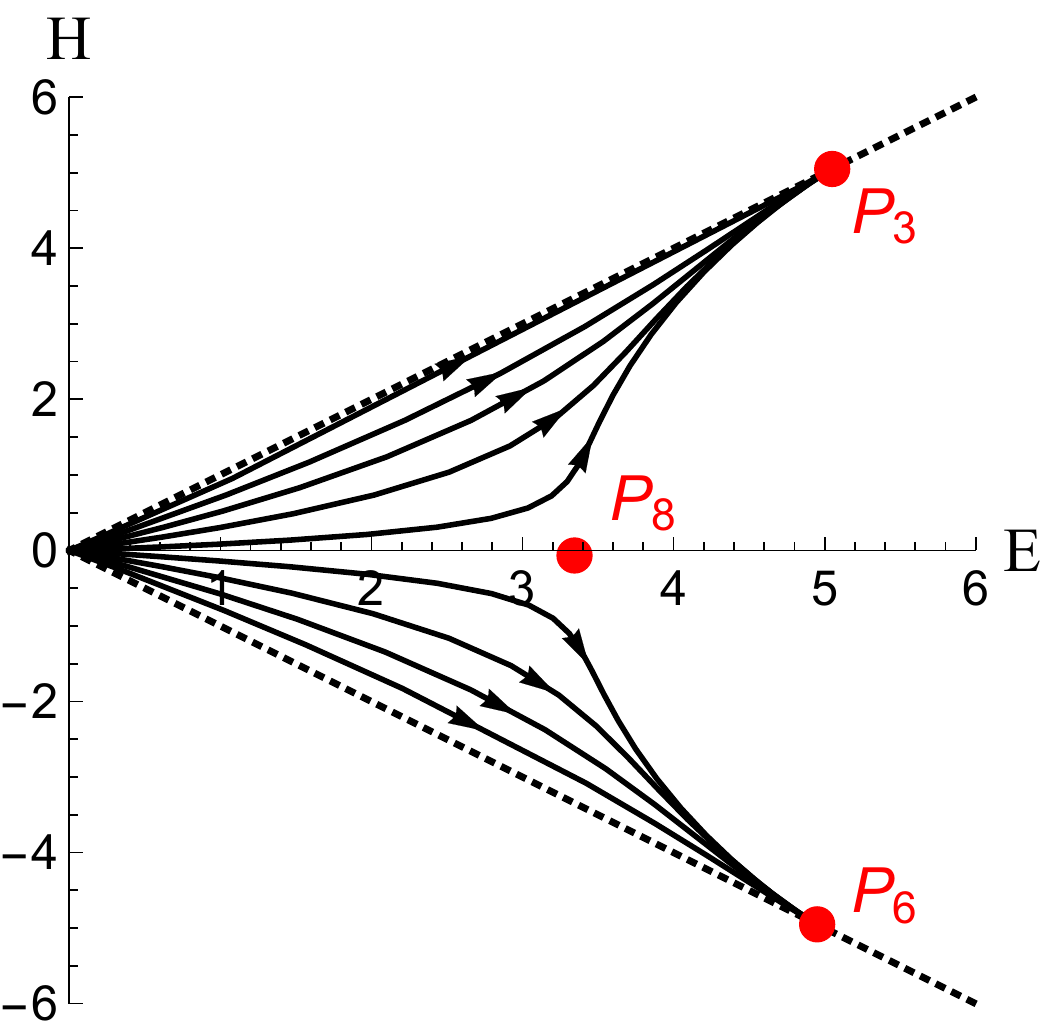}
\includegraphics[width=5cm]{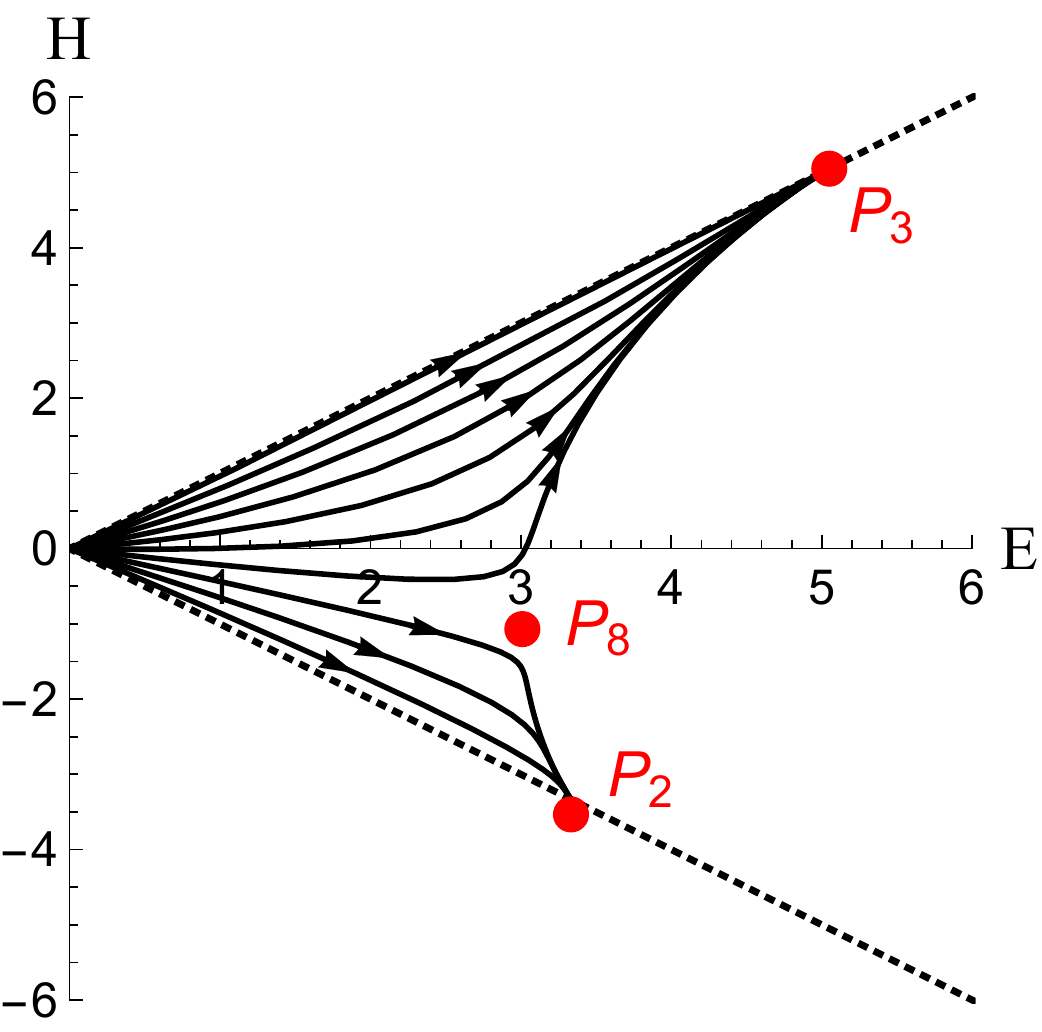}
\includegraphics[width=5cm]{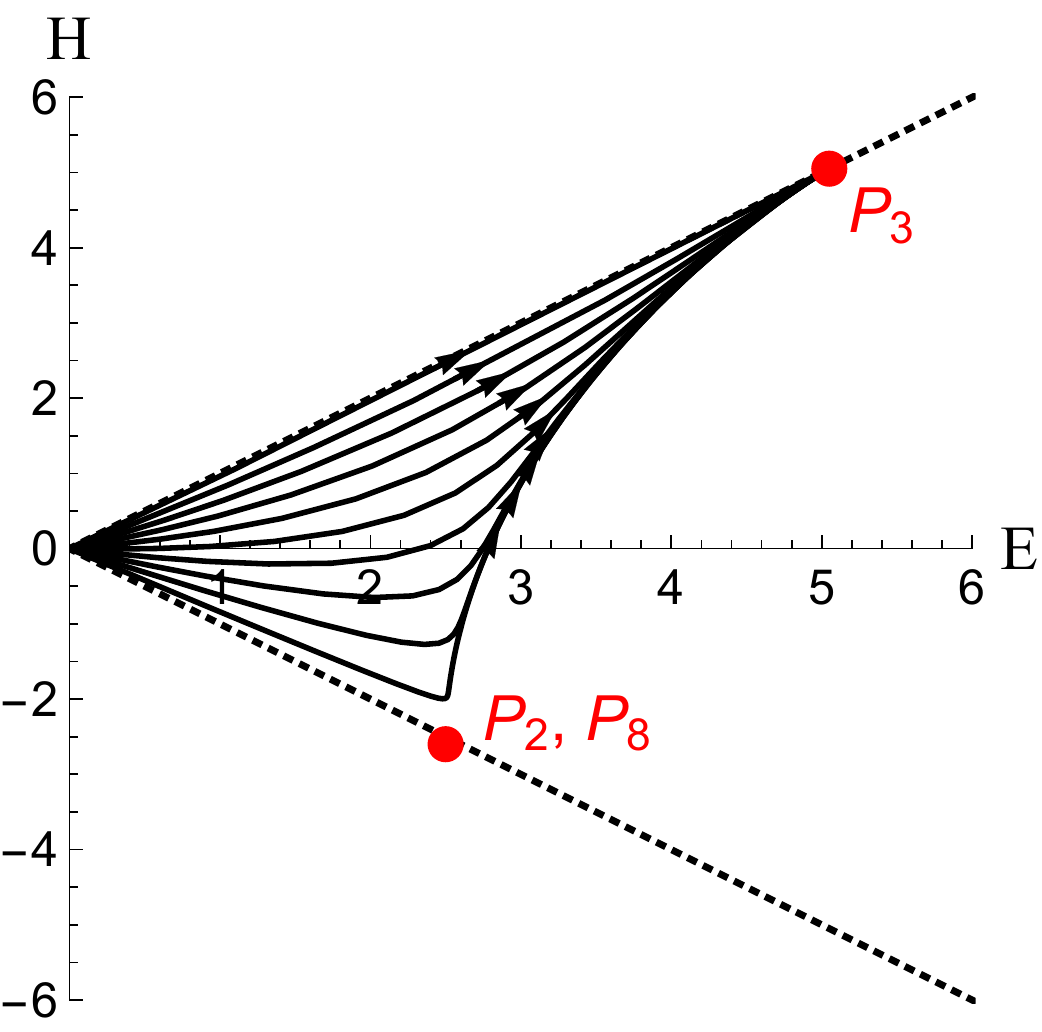}
\caption{From top to bottom: $\mu_L=0.10, 0.15, 0.20.$ The other parameters are set at $\mu_R = 0.1, \gamma_L = \gamma_R= 1, \mu^* = 0.2$.}
\label{fig:plotbroken}
\end{figure}
\end{center}

As mentioned in the introduction, in the case of a non-zero component $B_z$ the parity transformation $\mathcal{P}_{LR}$ is not a symmetry of the system anymore. The symmetry group is instead $P_{LR}{}^z$. We then expect the eigenmodes with opposite azimuthal wavenumbers to grow with different growth rates $\gamma_L$ and $\gamma_R$, which value is function of the symmetry-breaking component $B_z$. The same argument holds for all invariant monomials, hence leading to almost a doubling of parameters in our phenomenological description. Therefore, the effective Lagrangian in the weakly interacting case reads 
\ba\label{lagrangianESB}
\hsix \mathcal{L}(\bL,\bR)_{ESB} = \frac{1}{2}\, (\gamma_L\, \bL^2 + \gamma_R\, \bR^2) - \frac{1}{4}\, (\mu_L\, \bL^4 + \mu_R\, \bR^4) \nonumber\\- \frac{1}{2}\,\mu^*\, (\bL^2\, \bR^2) + \mathcal{O}(\bL^6) + \mathcal{O}(\bR^6) \, .
\ea
The time evolution equations for the energy and helicity densities read
\ba\label{ESBeqs}
\hhh d_t\, \text{H} = \gamma_L (\text{H} - \text{E}) + \gamma _R\, (\text{E} + \text{H}) + \mu_L\, (\text{E} - \text{H})^2 -  \mu _R \, (\text{E} + \text{H})^2 \nonumber ,\\[0.5em]
\hhh d_t\, \text{E} = - \text{E}^2  \left(2\, \mu ^*+\mu_L + \mu _R\right) + \text{E} \,\left(\gamma_L + 2\,\text{H}\,\mu_L - 2\,\text{H}\, \mu_R + \gamma_R\right) \nonumber \\ 
- \text{H}\, \left(\gamma_L + \text{H}\, \left(\mu_L + \mu_R - 2\, \mu^*\right) - \gamma_R\right)\, . \nonumber\\
\ea
The system (\ref{ESBeqs}) has eight fixed points $P_i = \{E^*{}_i, H^*{}_i\}$, $i = \{1, \cdots, 8\}$. Being a fourth-order approximation, we expect just four points to be physical, which interchange depending on were we are in parameter space. The catalogation of physical fixed points, together with their values and eigenvalues, are reported in the appendix B. Being (\ref{ESBeqs}) a generalization of the symmetric system (\ref{EHsys4}) we expect one fixed point to be the trivial solution $P_1 = \{E^*{}_1=0, H^*{}_1=0\}$, now with asymmetric eigenvalues
\ba
\Theta^{(1)}_1 = 
-\frac{\left| \gamma_L - \gamma_R \right| \left| \mu_L - \mu_R\right|}{\mu_L - \mu_R} + \gamma_L+\gamma_R\, ,\\
\Theta^{(1)}_2 =
+\frac{\left| \gamma_L - \gamma_R\right| \left| \mu_L - \mu_R\right|}{\mu_L - \mu_R} + \gamma_L + \gamma_R \, ;
\ea
a couple of fixed points at $H^* = E^*$; and an almost-symmetric fixed point at $H^* \approx 0$. As we broke the symmetry explicitly, the symmetric phase is never realized, hence leading to a fixed point where the the final helicity is of the order of symmetry breaking $\epsilon = \epsilon_\mu + \epsilon_\gamma$, $\epsilon_\mu = |\mu_L-\mu_R|$, $\epsilon_\gamma = |\gamma_L-\gamma_R|$.
It is depicted in Fig.\ref{fig:plotordered} the deviation from the symmetric phase in the initial model, by increasing the sole parameter $\epsilon_\mu$\footnote{The same qualitative picture can be obtained changing instead $\epsilon_\gamma$. Therefore, we hence present just one case for displaying purposes.}. As showed in the picture, as we break the symmetry the system saturates to a configuration with a final helicity of the order $\epsilon_\mu$, until helicity is maximized at a critical value.
In Fig.\ref{fig:plotbroken} we show instead how the broken phase in the initial model changes increasing $\epsilon_\mu$. As we explicitly break the symmetry, the fixed point values of energy and helicity at left and right points changes, while the saddle point moves towards one of the two helical solutions, and collapses on it at a critical value $\epsilon_\mu =(\epsilon_\mu)_{cr}$. In this regime, the sign of helicity at saturation is univocally determined despites the initial value. As in the higher-order case, an extended research in parameter space did not show any configuration with oscillations or limit cycles, while complex conjugated pair of eigenvalues are realized outside of the physical regime.

\section{Conclusions}
In the present work we investigated the contribution of higher-order terms to the helicity production described by an effective model based on a Lagrangian invariant under the symmetries of the system and a ground state that spontaneously breaks such a symmetry. Aim of the work was to understand whether higher non-linearities and the mechanism of mutual antagonism may alone explain the oscillatory patterns recently observed in simulations at low Prandtl numbers and high Hartmann numbers. Our higher-order analysis does indeed underline new peculiarities of the system, like solutions not maximining helicity and that lead to a physical scenario where saturation to a symmetric or broken case depends on the initial helicity of the perturbation. It does however not reproduce any oscillatory pattern, that is instead obtained just in the unphysical phase space. We then extended our analysis by investigating whether an explicit breaking of left-right symmetry due to a non-zero $z$-component of the perturbation may introduce oscillations. We found instead that the absence of oscillations is rather robust to symmetry breaking. 
It has moreover to be underlined that the oscillations unveiled in our work are due to complex eigenvalues of the fixed points, hence leading to oscillations both in energy and helicity, whether the oscillatory pattern observed in simulations shows only oscillations in helicity. Our results shows that an oscillating saturating state is unlikely to be reproduced by the interaction of chiral modes alone. One can argue that in order to reproduce an oscillating saturation state at high Hartmann numbers our effective Lagrangian must be modified to account for a dynamical coupling between magnetic and velocity perturbations. We plan to investigate this issue in a future work.

\acknowledgements
The research leading to these results has received funding from the European Community's Seventh Framework Programme (FP7/2007-2013) under grant agreement no. 269194. We thanks Norbert Weber and Frank Stefani for clarifying discussions on the numerical simulations.
\bibliographystyle{an}
\bibliography{ty}


\appendix
%
\onecolumn
\section{Spontaneuosly broken symmetry}
The fixed points in the case of spontaneuos symmetry breaking are:
\ba\label{ssbfp}
\hsix P_1 = \left\{\text{E}^*{}_1 = 0, \text{H}^*{}_1 = 0\right\}\, ,\\[1em]
%
\hsix P_2 = \left\{\text{E}^*{}_2= \frac{- \sqrt{4\, \gamma\, \left(6\, \lambda^* + \lambda \right) + \left(\mu^* + \mu\right)^2} - \mu^* - \mu}{2\, \left(6 \, \lambda^* + \lambda \right)}\, , \text{H}^*{}_2 = 0\right\} \, ,\\[1em]
%
\hsix P_3 = \left\{\text{E}^*{}_3 = \frac{\sqrt{4\, \gamma\, \left(6\, \lambda^* + \lambda \right) + \left(\mu^* + \mu \right)^2} - \mu^* - \mu}{2\, \left(6\, \lambda^* + \lambda \right)}\, , \text{H}^*{}_2 = 0\right\}\, ,\\[1em]
%
\hsix P_4 = \left\{\text{E}^*{}_4 = \frac{\sqrt{4\, \gamma\, \lambda + \mu^2} - \mu}{4\, \lambda}\,, \text{H}^*{}_4 =  - \frac{\sqrt{\frac{\left(\lambda - 2\, \lambda^*\right)\, \left(\mu  \left(\mu - \sqrt{4\, \gamma\,  \lambda + \mu^2}\right) + 2\, \gamma  \lambda\right)}{\lambda^2}}}{2\, \sqrt{2}\, \sqrt{\lambda - 2\, \lambda^*}}\right\}\, ,\\[1em]
%
\hsix P_5 = \left\{\text{E}^*{}_5 = \frac{\sqrt{4\, \gamma\, \lambda + \mu^2} - \mu }{4\, \lambda}\,, \text{H}^*{}_5 = \frac{\sqrt{\frac{\left(\lambda - 2\, \lambda^*\right)\, \left(\mu  \left(\mu - \sqrt{4\, \gamma\, \lambda + \mu^2}\right) + 2\, \gamma\, \lambda\right)}{\lambda^2}}}{2\, \sqrt{2}\, \sqrt{\lambda - 2\, \lambda^*}}\right\}\, ,\\[1em]
%
\hsix P_6 = \left\{\text{E}^*{}_6 = \frac{- \sqrt{4\, \gamma\, \lambda + \mu^2} - \mu}{4\, \lambda}\,, \text{H}^*{}_6 = - \frac{\sqrt{\frac{\left(\lambda - 2\, \lambda^*\right)\, \left(\mu\, \left(\sqrt{4\, \gamma\, \lambda +\mu^2} + \mu\, \right) + 2\, \gamma \, \lambda\right)}{\lambda^2}}}{2\, \sqrt{2}\, \sqrt{\lambda - 2\, \lambda^*}}\right\}\, ,\\[1em]
%
\hsix P_7 = \left\{\text{E}^*{}_7 = \frac{-\sqrt{4\, \gamma\,  \lambda + \mu^2} - \mu}{4 \, \lambda}\, , \text{H}^*{}_7 =  \frac{\sqrt{\frac{\left(\lambda - 2\, \lambda^*\right)\, \left(\mu\, \left(\sqrt{4\, \gamma\,  \lambda + \mu^2} + \mu \right) + 2\, \gamma  \, \lambda \right)}{\lambda^2}}}{2\, \sqrt{2}\, \sqrt{\lambda - 2\, \lambda^*}}\right\}\, ,\\[1em]
%
\hsix P_8 = \left\{\text{E}^*{}_8 = \frac{\mu^* - \mu}{2\, \left(\lambda - 2\, \lambda^*\right)}\, , \text{H}^*{}_8 = - \frac{\sqrt{\frac{4\, \gamma \, \left(\lambda - 2\, \lambda^*\right)^2 + \left(\mu - \mu^*\right)\, \left(2\, \lambda^*\, \left(\mu^* - 5\, \mu \right) + \lambda\,  \left(3\, \mu^* + \mu\, \right)\right)}{\left(\lambda - 2\, \lambda^*\right)^2}}}{2\, \sqrt{\lambda - 2\, \lambda^*}}\right\}\, ,\\[1em]
%
\hsix P_9 = \left\{\text{E}^*{}_9 = \frac{\mu^* - \mu}{2\, \left(\lambda - 2 \, \lambda^*\right)}\, , \text{H}^*{}_9 = \frac{\sqrt{\frac{4\, \gamma  \left(\lambda - 2\, \lambda^*\right)^2 + \left(\mu - \mu ^*\right)\, \left(2\, \lambda^*\, \left(\mu ^* - 5\, \mu\, \right) + \lambda \, \left(3\,  \mu^*  + \mu \right)\right)}{\left(\lambda - 2 \, \lambda^*\right)^2}}}{2\, \sqrt{\lambda - 2\, \lambda^*}}\right\}\, .
\ea
\newline
Their eigenvalues read:
\ba\label{ssbfpeigenv}
\hsix \Theta^{(1)}{}_1 = \Theta^{(1)}{}_2 = 2\, \gamma \, ,\\[1em]
%
\hsix \Theta^{(2)}_1 = - \frac{\left(\mu^*+\mu \right)\, \left(\sqrt{4\, \gamma\,  \left(6\, \lambda^* + \lambda \right) + \left(\mu^* + \mu \right)^2} + \mu^* + \mu \right) + 4\, \gamma\,  \left(6\, \lambda^* + \lambda \right)}{6\, \lambda^* + \lambda}\, ,\\[1em]
\hsix \Theta^{(2)}_2 = -\Big\{\left(2\, \lambda^*\, \left(\mu^* - 5\, \mu\right) + \lambda\, \left(3\, \mu^* + \mu \right)\right)\, \Big(\sqrt{4 \, \gamma\,  \left(6\, \lambda^* + \lambda\right) + \left(\mu^* + \mu\right)^2} \nonumber \\
+ \mu^* + \mu \Big) + 4 \, \gamma\, \left(\lambda - 2 \, \lambda^*\right)\, \left(6\, \lambda^* + \lambda\right)\Big\}\, \frac{1}{\left(6\, \lambda^* + \lambda\right)^2}\, ,\\
%
\hsix \Theta^{(3)}{}_1 = \frac{\left(\mu^*+\mu \right)\, \left(\sqrt{4\, \gamma\, \left(6\, \lambda^* + \lambda \right) + \left(\mu^* + \mu \right)^2} - \mu^* - \mu \right) - 4\, \gamma\, \left(6\, \lambda^* + \lambda \right)}{6\, \lambda^* + \lambda}\, ,\\[1em]
\hsix \Theta^{(3)}{}_2 = \Big\{\left(2\, \lambda^*\, \left(\mu^* - 5\, \mu\right) + \lambda\, \left(3\, \mu^* + \mu \right)\right)\, \Big(\sqrt{4\, \gamma\, \left(6\, \lambda^* + \lambda\right) + \left(\mu^* + \mu\right)^2} \nonumber \\
- \mu^* - \mu\Big) - 4\, \gamma \, \left(\lambda - 2\, \lambda^*\right)\, \left(6\, \lambda^* + \lambda\, \right)\Big\}\, \frac{1}{\left(6\, \lambda^* + \lambda \right)^2}\, ,\\[2em]
%
\hsix \Theta^{(4)}{}_1 = \Theta^{(5)}{}_1 = \Theta^{(4,5)}_A + \Theta^{(4,5)}_B\, ,\\[1em]
\hsix \Theta^{(4)}{}_2 = \Theta^{(5)}{}_2 = \Theta^{(4,5)}_A - \Theta^{(4,5)}_B\, ,\\[1em]
\hsix \Theta^{(4,5)}_A = \frac{\left(2\, \lambda^*\, \mu + \lambda\,   \left(\mu - \mu^*\right)\right)\, \left(\sqrt{4\, \gamma\, \lambda + \mu^2} - \mu \right) - 2\, \gamma\, \lambda\, \left(2\, \lambda^*+  \lambda \right)}{2\, \lambda^2}\, ,\\[1em]
\hsix \Theta^{(4,5)}_B = \frac{1}{2\, \lambda^2\, \sqrt{\lambda - 2\, \lambda^*}}\, \sqrt{2}\, \Big\{\left(\lambda - 2\, \lambda^*\right)\, \left(2\, \gamma^2\, \lambda^2\, \left(3\, \lambda - 2\, \lambda^*\right)^2 - \mu\, \left(\lambda\, \left(\mu^* + \mu \right) \right. \right. \nonumber \\
\hsix \left. \left. - 2\, \lambda^*\, \mu \right)^2 \, \left(\sqrt{4\, \gamma\,  \lambda + \mu^2} - \mu \right) - 2\, \gamma\,  \lambda\,  \left(\lambda\, \left(\mu^* + \mu \right) - 2\, \lambda^*\, \mu \right) \,\left(- 2\, \lambda^* \, \sqrt{4\, \gamma\, \lambda +\mu^2} \right.\right. \nonumber \\
\left. \left. + 3\, \lambda\, \sqrt{4\, \gamma\, \lambda + \mu^2} + 4\, \lambda^*\, \mu - \lambda\, \mu^* - 4\, \lambda\, \mu\right)\right)\Big\}^{\frac{1}{2}}\, ,\\[2em]
%
\hsix \Theta^{(6)}{}_1 = \Theta^{(7)}{}_1 = \Theta^{(6,7)}_A + \Theta^{(6,7)}_B\, ,\\[1em]
\hsix \Theta^{(6)}{}_2 = \Theta^{(7)}{}_2 = \Theta^{(6,7)}_A - \Theta^{(6,7)}_B\, ,\\[1em]
\hsix \Theta^{(6,7)}_A = \frac{\left(2\, \lambda^* \, \mu + \lambda\,  \left(\mu - \mu^*\right)\right)\, \left(\sqrt{4\, \gamma \, \lambda + \mu^2} + \mu \right) + 2\, \gamma\, \lambda \, \left(2\, \lambda^* + \lambda \right)}{2\, \lambda^2}\, , \\[1em]
\hsix \Theta^{(6,7)}_B = \frac{1}{2\, \lambda^2\, \sqrt{\lambda - 2\, \lambda^*}}\,\Big\{\left(\lambda - 2\, \lambda^*\right)\, \left(2\, \gamma^2\, \lambda^2\, \left(3\, \lambda - 2\, \lambda^*\right)^2 + \mu\, \left(\lambda\, \left(\mu^* + \mu\right)
\right. \right. \nonumber \\ \left. \left.
\hsix - 2\, \lambda^*\, \mu\right)^2\, \left(\sqrt{4\, \gamma\, \lambda + \mu^2} + \mu\right) + 2\, \gamma\, \lambda\,  \left(\lambda\,  \left(\mu^* + \mu\right) - 2\, \lambda^*\, \mu \right)\, \left(\lambda \, \Big(3\, \sqrt{4\, \gamma \, \lambda + \mu^2} 
\right. \right. \nonumber \\ 
\hsix \left. \left. + \mu^* + 4\, \mu \Big) - 2\, \lambda^*\, \left(\sqrt{4\, \gamma \, \lambda + \mu ^2} + 2\, \mu\right)\right)\right)\Big\}^{\frac{1}{2}}\, ,\\[2em]
%
\hsix \Theta^{(8)}{}_1 = \Theta^{(9)}{}_1 = \Theta^{(8,9)}_A + \Theta^{(8,9)}_B\,  ,\\[1em]
\hsix \Theta^{(8)}{}_2 = \Theta^{(9)}{}_2 = \Theta^{(8,9)}_A - \Theta^{(8,9)}_B\, , \\[1em]
\hsix \Theta^{(8,9)}_A = \frac{- 4\, \gamma \, \left(\lambda - 2\, \lambda^*\right)^5 - \left(\lambda - 2\, \lambda^*\right)^3\, \left(\mu - \mu^*\right)\, \left(\lambda\, \left(2\, \mu^* + \mu \right) - 6\, \lambda^*\, \mu \right)}{\left(\lambda - 2\, \lambda^*\right)^5}\, , \\[1em]
\hsix \Theta^{(8,9)}_B = \frac{1}{\left(\lambda - 2\, \lambda^*\right)^5}\, \Big\{\left(\lambda - 2\, \lambda^*\right)^6\, \left(\mu - \mu^*\right) \, \left(4\, \gamma\,  \left(\lambda - 2\, \lambda^*\right)^2\, \left(- 2\, \lambda^*\, \mu^* + 6\, \lambda^*\, \mu \right.\right. \nonumber \\ 
\left. \left. - 3\, \lambda  \mu^* + \lambda\, \mu\right) + \left(\mu - \mu^*\right)\, \left(8\, \lambda^*\, \left(2\, \lambda^* + 5\, \lambda \right)\, \mu^*\, \mu - 8\, \lambda \, \left(\lambda^* + \lambda\right)\, \left(\mu^*\right)^2 \right. \right. \nonumber \\ \left. \left.
+ \left(\lambda^2 - 4\, \lambda^*\, \lambda - 44\, \left(\lambda^*\right)^2\right)\, \mu^2\right)\right)\Big\}^{\frac{1}{2}}\, .
\ea
\section{Explicitly broken symmetry}
The fixed points in the case of explicit symmetry breaking are:
\ba
\hsix P_1 = \{\text{E}^*{}_1 = 0, \text{H}^*{}_1 = 0\} \, ,\\[1em]
\hsix P_2 = \left\{\text{E}^*{}_2 = \frac{\gamma_L}{2\, \mu _L}, \text{H}^*{}_2 = \frac{\gamma_L\, \mu _R - \mu_L \left(\left| \gamma_L - \gamma _R \right| + \gamma_R\right)}{2\, \mu_L\, \left(\mu_L - \mu_R\right)}\right\} \, ,\\[1em]
\hsix P_3 = \left\{\text{E}^*{}_3 = \frac{\gamma_R}{2\, \mu _R}, \text{H}^*{}_3 = \frac{\mu_L \,\gamma_R - \mu_R\, \left(\left| \gamma_L - \gamma_R \right| + \gamma_L\right)}{2\,\mu_R\, \left(\mu_L - \mu_R\right)} \right\} \, ,\\[1em]
\hsix P_4 = \left\{\text{E}^*{}_4 = \frac{- \mu^*\, \left(\gamma_L + \gamma_R\right) + \mu_L\, \gamma_R + \gamma_L\, \mu_R}{2\, \mu_L\, \mu_R - 2\, \left(\mu^*\right)^2}, \right. \nonumber \\ 
\hhh \left. \text{H}^*{}_4 = - \left(\left| \frac{\left(\gamma_L + \gamma_R\right)\, \left(\mu^*\right)^2 - 2\,
   \left(\gamma_R\, \mu _L + \gamma_L\, \mu_R\right)\, \mu^* + \left(\gamma_L + \gamma _R\right)\, \mu _L\, \mu _R}{\left(\mu^*\right)^2 - \mu_L\, \mu _R}\right| +\gamma_L  \right. \right. \nonumber \\ \left. \left. 
\hhh+\frac{\left(\mu_L + \mu_R\right)\,  \left(\mu^*\, \left(\gamma_L + \gamma_R\right) + \mu_L\, \left( - \gamma _R\right) - \gamma_L\, \mu _R\right)}{\mu_L\, \mu_R - \left(\mu^*\right)^2} + \gamma_R \right)\, \frac{1}{2\, \left(\mu _L - \mu _R\right)}\right\}\, ,\\[1em]
\hsix P_5 = \left\{\text{E}^*{}_5 = 0,\, \text{H}^*{}_5 = \frac{\gamma_L + \gamma_R}{\mu_R -\mu _L}\right\} \, ,\\[1em]
\hsix P_6 = \left\{\text{E}^*{}_6 = \frac{\gamma_L}{2\, \mu_L}, \text{H}^*{}_6 = \frac{\mu_L \left| \gamma_L - \gamma_R \right| - \mu_L\, \gamma_R + \gamma_L\,\mu_R}{2\, \mu_L^2 - 2\, \mu_L\, \mu_R}\right\} \, ,\\[1em]
\hsix P_7 = \left\{\text{E}^*{}_7 = \frac{\gamma_R}{2\, \mu_R}, \text{H}^*{}_7 = \frac{\mu_R \, \left| \gamma_L - \gamma_R\, \right| + \mu_L\, \gamma_R - \gamma_L\, \mu_R}{2\, \mu_L\, \mu_R - 2\, \mu_R^2}\right\} \, ,\\[1em]
\hsix P_8 =\left\{\text{E}^*{}_8 = \frac{- \mu^*\, \left(\gamma_L + \gamma_R\right) + \mu_L\, \gamma_R + \gamma_L\, \mu_R}{2\, \mu_L\, \mu_R - 2\,\left(\mu^*\right)^2}, \right. \nonumber \\ 
\hhh \left. \text{H}^*{}_8 = - \left(-\left| \frac{\left(\gamma_L + \gamma_R\right)\, \left(\mu^*\right)^2 - 2\, \left(\gamma_R\, \mu_L+\gamma _L\, \mu _R\right)\, \mu^* + \left(\gamma_L + \gamma_R\right)\, \mu_L\, \mu_R}{\left(\mu^*\right)^2 - \mu_L\, \mu _R}\right| +\gamma_L \right. \right. \nonumber \\
\hhh \left. \left. +\frac{\left(\mu_L + \mu_R\right)\, \left(\mu^* \left(\gamma_L + \gamma_R\right) + \mu_L\, \left( - \gamma_R\right) - \gamma_L\, \mu _R\right)}{\mu_L\, \mu_R - \left(\mu^*\right)^2} + \gamma_R\right)\frac{1}{2\, \left(\mu_L - \mu _R\right)}\right\} \, .
\ea

\end{document}